\documentclass[letterpaper]{article} 
\usepackage{aaai2026}  
\usepackage{times}  
\usepackage{helvet}  
\usepackage{courier}  
\usepackage[hyphens]{url}  
\usepackage{graphicx} 
\usepackage{natbib}  
\usepackage{caption} 
\usepackage{subcaption}
\usepackage{graphicx}
\usepackage{amsmath}
\usepackage{soul}
\usepackage{xcolor}  
\usepackage{colortbl}
\usepackage{multirow}
\usepackage{booktabs}
\usepackage{algorithm}
\usepackage{algorithmic}

\usepackage{xcolor}

\usepackage{newfloat}
\usepackage{listings}
\DeclareCaptionStyle{ruled}{labelfont=normalfont,labelsep=colon,strut=off} 
\floatstyle{ruled}
\newfloat{listing}{tb}{lst}{}
\floatname{listing}{Listing}
\title{From Scroll to Sale: Exploring the Impact of Interaction Type\\ and Device Price on TikTok Advertisements}
\author {
    Nazanin Sabri\textsuperscript{\rm 1},
    Cat Mai\textsuperscript{\rm 2}, 
    Haodi Zou\textsuperscript{\rm 1},
    Isha Varada\textsuperscript{\rm 1},
    Damon McCoy\textsuperscript{\rm 2},
    Deepak Kumar\textsuperscript{\rm 1},
    Kristen Vaccaro\textsuperscript{\rm 1}
}
\affiliations {
    \textsuperscript{\rm 1}University of California San Diego\\
    \textsuperscript{\rm 2}New York University\\
    \{nsabri, h4zou, ivarada, d1kumar, kvaccaro\}@ucsd.edu, 
    \{cat.mai, mccoy\}@nyu.edu
}

\usepackage{booktabs}
\usepackage{multirow}
\usepackage[table]{xcolor}

\begin{document}

\maketitle

\begin{abstract}
Companies and brands increasingly use dynamic pricing, including targeting social media ads to users based on their income.  
In this work we audit TikTok's feed using 56 automated accounts, which collect data on over 80,000 videos, across two studies. 
We test the impact of device price on ad load and ad types, using 12 phones of low (\$0-\$250), medium (\$400-\$650), and high (\$750-\$1,000+) price as a proxy for income. We also test the impact of interaction type (i.e., like, comment, share), age, and gender 
on the frequency and content of ads.  
Overall, 
the ad load was 
29.4\%, but liking and sharing content increased ad load significantly. We also found that as accounts spend more time on TikTok, the ad load steadily increases. We found some evidence that device price impacts both ad load and content --- more expensive devices were targeted with fewer ads, while the least expensive devices received more discounts. 
\end{abstract}


\section{Introduction}

A recent TikTok video went viral after accusing retailers of changing in store prices in real time~\cite{yahoo2025walmarttiktok}.
But online this kind of dynamic pricing is already entrenched. 
Airlines and hotels, 
for example, have used dynamic pricing to manage demand in online reservations for decades~\cite{mcafee2008price, ftc_surveillance, wsj_orbitz}.  
Companies argue that dynamic pricing allows 
motivated consumers to seek out the best deals
~\cite{patterson2025pricediscrimination, mcafee2008price}. However, this 
creates a risk that low-income shoppers 
will face higher prices or lower quality goods when sellers believe they have less power to walk away~\cite{wsj_internet}. 

In this study, we examine the extent to which TikTok advertisements target users differentially based on income. Despite the large volume of work studying targeted advertising, including its impacts on job opportunities, political information, and the reach of scams, the focus has remained primarily on demographics like age, race, and gender~\cite{10.1145/3442381.3450077, ali_problematic_2023, 10.1145/3366423.3380109}. 
Short-form video platforms have also rarely been the focus of advertising audits~\cite{10.1145/3589335.3651260, min2025impact}, despite their increasing popularity~\cite{10.1145/3485447.3512102}. The only 
study of targeted advertising on TikTok 
examined the impact of other demographics, specifically age and gender, on the kind of ads that users are shown~\cite{10.1145/3589335.3651260}. Our study contributes the first measurement of the 
impact of income on the volume, content, and prices of targeted advertisements.

Before studying the impact of income and other user demographics, however, it is necessary to understand the effects of engagement on ad load. Both on TikTok and other platforms, users' engagement with their feeds 
have been shown to shape the content they subsequently encounter
~\cite{germano2026ranking, shin2024algorithms}. This phenomenon, however, has 
not been studied for advertisements. This leads to our first research question: 
\textbf{RQ1}: How does interaction type (i.e., scrolling, liking, commenting, sharing) affect the number and content of ads?


Next, we focus on the impact of income. Marketing firms have noted the value of targeting advertisements based on income or wealth~\cite{john2023exploration}, so we examine the extent to which that is currently practiced in targeted advertising campaigns on TikTok.
While TikTok offers an ``About This Ad'' explanation for each ad
~\cite{tiktok2022aboutthisad}, 
prior work has shown that  up to 40\% of TikTok's explanations are incorrect
~\cite{mousavi2024auditing}. Instead, we opt to directly measure the impact of income, which we experimentally vary using device price as a proxy, on the number and type of ads shown. This forms our second research question: 
\textbf{RQ2}: How does income affect the number and content of ads?

Finally, age and gender are among the most commonly exploited demographic signals in targeted advertising~\cite{okoroafo2010gender}. Prior work on other online spaces has shown gender disparities in ad targeting~\cite{10.1145/3442381.3450077, lambrecht_algorithmic_2019, ali_problematic_2023}. This 
evidence from similar platforms motivates our last research question: 
\textbf{RQ3}: How do 
age and gender 
affect the number and content of ads?

To answer these questions, we performed two paired 
studies using an automated system to view, interact with, and record TikTok content. Our first study tested four interaction types (scroll, like, comment, and share) to identify whether any interactions increased the ad load. Using 32~automated accounts, we collected data on 42,819~videos. 
Overall, the \emph{ad load} was 29.4\%. However, some more active forms of engagement (liking and sharing) dramatically increased ad load, up to 37.6\% of the videos encountered. 
Surprisingly, commenting did not lead to the same effect --- in fact, the platform seemed to treat frequently commenting very differently from all other forms of interaction. 

Our second study tested the impact of device price (as a proxy for income) and demographics (age and gender) on the number and type of ads shown, to answer \textbf{RQ2} and \textbf{RQ3}. To test device price, we used 12 phones, with prices ranging from \$80--\$1,000. To test demographics, we created accounts with two ages (26 and 46) and two genders, suggested through the choice of account name. 
The 24 automated accounts in Study 2 more closely emulated real user interactions, based on interaction patterns obtained from TikTok user data donations~\cite{zannettou2023leveraging}, eventually collecting data on 37,350 videos. 
We found that our most expensive phone had the lowest ad load, but the least expensive device received the largest discounts.  

These findings suggest important implications for the ad ecosystem, for example, that users who do not have the means or desire to purchase more expensive devices may
face a higher advertisement burden, with ads taking up more
of their time and attention. However, the importance of interactions, relative to demographic attributes, suggests ways for everyday users to exert more power over what they see. 






\section{Related Work}

Algorithmic feeds --- like TikTok's ``For You'' page or Instagram's ``Explore'' feed --- are 
integral parts of modern social media. 
These feeds rely on a content recommendation algorithm that match users with content they might enjoy seeing. However, these algorithms are usually not made visible nor transparent to end-users~\cite{10.1145/3726302.3730293, edelson_comparative_2025}. Given the prevalence of these platforms and the high impact they can have, researchers~\cite{10.1145/3449148} and regulators~\cite{ec2023audit, blumenthal2023letter} have become increasingly interested in auditing social media platforms. 



\subsubsection{Social Media Audits}

Algorithm audits, defined as ``methods that systematically probe algorithms and observe their output to draw conclusions about their inner workings''~\cite{10.1145/3555625}
, have emerged as means of evaluating algorithms' problematic behavior.
Sandvig et al. \cite{sandvig_auditing_2014} established foundational frameworks for detecting algorithmic discrimination, 
proposing systematic approaches to reveal hidden biases in automated decision-making systems.

\subsubsection{Audits of Ad Delivery}

Audits have been performed on a variety of functions of social media platforms; one 
important function is targeted advertising.
Given the black box nature of ad targeting systems, 
they can result in unintended outcomes.
Studies have found that the final audience of an ad can significantly differ from what the advertiser originally intended after ad delivery optimization~\cite{lambrecht_algorithmic_2019,asplund_auditing_2020,ali_problematic_2023}.
 
Industry white papers
have reported that users 
prefer relevant ads to generic ones 
~\cite{mckinsey_personalization}, though other researchers have found 
that users find targeted ads creepy~\cite{reitinger2024does, ur_smart_2012, wei_what_2020} and view the 
behavioral tracking required for 
ad delivery optimization
unfavorably~\cite{reitinger2024does}.
Social media users find that current transparency mechanisms are unhelpful in understanding their ad composition~\cite{ur_smart_2012, mousavi2024auditing} and attempts to filter out unwanted ads are ineffective~\cite{kaplan-2026-tiktok}.

Prior work has examined how ad delivery can be influenced by user's demographics such as gender, race, or sexual orientation~\cite{ali_problematic_2023,kofman2019facebook, asplund_auditing_2020, gak_distressing_2022}.
In 2012, investigations revealed that the travel website Orbitz directed Mac OS users toward higher-priced hotels by ranking these properties higher in search results~\cite{wsj_orbitz}.
There is also an emerging line of work estimating how valuable users are and the resulting ads that they see~\cite{cabanas_unveiling_2018, mai2025youtubescamads}.
Mai et al.\ showed that private users, arguably less valuable to advertisers, received a much higher ad load than when their less private counterparts.
To the best of our knowledge, no prior work has investigated the relationship between both user's device price and behavior and the resulting ads they get on TikTok.



\subsubsection{Audits of TikTok}

Audits of TikTok have utilized a variety of methods such as sockpuppet accounts~\cite{mousavi2024auditing}, data donations~\cite{10.1145/3630744.3659830}, and analysis of network traffic~\cite{kaplan-2026-tiktok}.
Most studies have focused on video recommendations, though some have considered elements like 
auto complete search suggestions, where misinformation is common~\cite{gunasekara5580515auditing}.

Prior audits of TikTok have found that age impacts advertisements, showing that ads shown to two adult accounts (18 years old) were more similar to one another than those shown to child accounts (13 years old)~\cite{10.1145/3589335.3651260}. 
However, the majority (83\%) of videos shown on 
a child's recommendation feed 
are not child directed~\cite{figueira2025kids} and 
the mechanisms meant to limit exposure to harmful content on youth accounts are only minimally effective, with little difference 
between adult and youth accounts~\cite{xue2025towards}.  

Researchers have also studied TikTok's explanations for recommended videos, 
finding that a large number of explanations provided for why a content was shown was incompatible with the behavior of the automated 
account~\cite{mousavi2024auditing}. For example, an account that never commented, yet, was provided the explanation ``You commented on similar videos''. 

Methodologically closely related to our work is~\cite{10.1145/3485447.3512102} which investigated the impacts of interaction type on content recommendation. In order, the most influential interactions were follow, like, and watch rate
~\cite{10.1145/3485447.3512102}. While this work focused on content recommendations, we extend the analysis to advertisements and add additional interaction types. 

\renewcommand{\arraystretch}{1.2}

\begin{table}
    \centering
    \begin{tabular}{l l l}
    \multicolumn{3}{l}{\textit{Experimental Conditions}}\\
     \textbf{Device Price} &  \textbf{Interaction} & \textbf{Demographics} \\ [.2em]
    \toprule
    \multicolumn{3}{l}{\textbf{Study 1} -- 42,819 videos and 12,608 ads} \\ [.2em]
   \midrule

         \multirow[t]{4}{1.9cm}{\textcolor{gray}{\$80$^\dagger$} \\\textit{\textcolor{gray}{Pixel 3a}}} & \multirow[t]{4}{1.75cm}{Scroll \\[.2em] Like \\[.2em] Comment \\[.2em] Share} & \multirow[t]{6}{3.75cm}{\textcolor{gray}{1980 Michael or Jennifer}\\ [.2em]
        \textcolor{gray}{2000 Jacob or Emily$^\dagger$}}\\
 & & \\
 & & \\
 & & \\ 
[1.2em]
\toprule

\multicolumn{3}{l}{\textbf{Study 2} -- 37,350 videos and 11,647 ads} \\[.2em]
\midrule
 \multirow[t]{6}{1.9cm}{\$80 \\\textit{Pixel 3a} \\[.5em] \$499 \\ \textit{Pixel 8 }\\[.5em] \$924 \\ \textit{Galaxy S26+}} & \multirow[t]{6}{1.75cm}{\textcolor{gray}{All$^\dagger$}} & \multirow[t]{6}{3.75cm}{1980 Michael or Jennifer \\ [.3em] 2000 Jacob or Emily}\\
 & & \\
 & & \\
 & & \\
 & & \\
 & & \\
\bottomrule
    \end{tabular}
    \caption{Experimental design overview. Study 1 tested the impact of different interaction types, while Study 2 tested the impact of device price (i.e., wealth) and demographics. $\dagger$While all settings were required to collect data (e.g., accounts cannot be created without an age and name), results use data only from 
    the respective studies. 
    }
    \label{tab:summary_of_studies}
\end{table}


\section{Experimental Design}
\label{section:exp_design}

In order to understand the factors that impact the delivery of advertisements on TikTok, we conducted an audit of the platform through two paired studies. In our first study, we investigated how varying types of user engagement with recommendation feeds impacted the number and content\footnote{
Content is operationalized as visual representations, brands, and price and is described in detail in Dependent Variables.} of ads they received. In our second study, we explored how advertising exposure differed across low- and high-income users, revealing potential disparities in nature and price of ads delivered to different socioeconomic groups. Table \ref{tab:summary_of_studies} summarizes the experimental conditions. 

The device price condition was selected as a proxy for income, because prior work has shown that ad targeting based on income could disadvantage lower-income individuals, who may have less flexibility to seek out better alternatives if higher prices are shown to them~\cite{patterson2025pricediscrimination}. Demographics such as age and gender have long been used by marketers~\cite{okoroafo2010gender}, so accounts were created with different names and dates of birth 
to assess how these identity markers shape the ads users receive. Finally, the interaction type condition was examined, as prior work has shown users' engagement with algorithmic feeds to influence the content they subsequently encounter~\cite{germano2026ranking}.



To perform 
the audit, we built a system to 
automatically scroll, interact with, and record content on TikTok's recommendation feed. 
After we (1) manually created accounts with matched ages and gender, the automated tool for each account 
(2) 
scrolled TikTok's recommendation feed, (3) performed different kinds of interactions, and (4) collected and stored screenshots of all videos. An analysis pipeline then 
(5) identified ads and (6) analyzed their content. 
Data collection for each setting was duplicated, to control for randomness in TikTok's algorithm, 
i.e., we created two accounts with the exact same name, age, and interaction behavior. Concurrent data collection across over a dozen devices 
controlled for marketing campaigns over time. 
The data analyzed throughout this study was collected between March 12 and April 25, 2026. In total, 80,169 TikTok videos were encountered, 
24,255 (30\%) of which were advertisements. 

While many audits using sock puppet accounts are not considered human subjects research, we obtained IRB approval for this study. One experimental condition (the like and comment interaction types) 
created opportunities for interactions with human users of the platform, 
including both 
perceptions by content creators and the fact that some of our automatically generated comments received likes or replies from other users. To minimize potential deception or harm, 
our comments consisted only of randomly selected emojis. 

\subsection{Independent Variables}

We modified 
interaction type, age, gender (proxied through name), and income (proxied through device price) to answer our research questions. Study 1 examined the effects of interaction behavior on ad load\footnote{Ad load is operationalized as the percentage of ads.}, while Study 2 explored the effects of income, age and gender. 

\subsubsection{Interaction Behaviors}

How users interact with their feeds on TikTok -- through watch time, likes, comments, and shares -- can strongly influence the content 
they are recommended, as these engagement signals are used to infer user interests and preferences~\cite{10.1145/3485447.3512102}. However, these were 
measurements of content recommendations, so our study extended 
the measurement to advertising. 

Five interaction types were examined: scroll, like, comment, share, and all interactions combined. In the first study, interactions were isolated by performing only one interaction per account to identify their individual effects. The second study combined interactions to better simulate more realistic user behavior. 

\vspace{3pt}
\noindent \textit{Interaction Times Throughout The Day} Each account collected TikTok content for 2:15–3 hours, aligning with estimates of average social media use~\cite{soax_social_media_time_2025}. 
 Usage was distributed across multiple sessions, rather than a single continuous period, to reflect habitual checking behaviors throughout the day. This type of behavior, where users come back to the application throughout the day, has been shown in prior data donations of the platform~\cite{yang2025studying}. Usage was spread across five sessions: a 45–60 minute period in the morning, three randomly timed 15–20 minute sessions during the day, and a 45–60 minute period at night. The time of day and duration of the five interactions were selected at random at the beginning of each day. 

\vspace{3pt}
\noindent \textit{Interaction Probability} In every experiment there was 
a 10\% chance of interacting with each video. The type of interaction depended 
on the study and experimental condition. 

\vspace{3pt}
\noindent \textit{Interaction Types} 
Study 1 tested interaction types independently (though all accounts populated new content by scrolling). Scrolling served as the baseline condition, in which content was only viewed and no engagement with the feed occurred. View time was selected at random ranging from 1 to 25 seconds. If a video was shorter than the length selected it would loop until the program scrolled past it. Re-watching content is not uncommon behaviors for users~\cite{zebracat_tiktok_statistics}.


The other interactions were: liking, commenting, and sharing content. Follow was initially included within our experimental conditions, but was later removed from the study. This is due to a follow-ban placed on the device running this interaction. Manual attempts at following users on the original account performing this interaction, as well as new ones created, were unsuccessful.

Study 2 aimed to replicate more realistic interaction behavior. All accounts 
liked 93\% of the time, commented 
5\% of the time, and shared 2\% of the time. These percentages were 
based on real TikTok usage statistics reported in prior work using data donations~\cite{zannettou2023leveraging}.




\subsubsection{Income} 

Income could be used by advertisers to tailor campaigns toward users with higher purchasing power. It might also be used to present higher prices or lower-quality goods to lower income users who may have limited flexibility in their options~\cite{patterson2025pricediscrimination}. 


To understand how income might be used by marketers to target ads, we evaluated the impact of device price, as a proxy for income on ad targeting. TikTok's advertising platform includes device price as one of the parameters advertisers can adjust, supporting the idea 
that this variable may be used for ad targeting.


Devices from three price ranges tested the effects of price on ads received. These price ranges were: \$0-\$250, \$400-\$650, and \$750-\$1,000+, which according to the platform have approximately equivalent audience size estimates. 
Large price differences 
(\$150) 
were placed between the ranges to allow for differentiation. Appendix A provides audience counts for each group.


\subsubsection{Age} 
Marketers have emphasized the importance of tailoring marketing campaigns to specific age~\cite{amazon_generational_marketing, fromm2018marketing}. To investigate the potential effects of age on ad targeting on TikTok, we tested two age values within our experiments: 26 and 46. 

Age 26 was selected because it is within the largest adult age group (i.e., 25-34) on TikTok in 2025~\cite{explodingtopics_tiktok_demographics_2026}. The second age, 
46, was then selected because it 
enabled comparison between ads served to younger and older adults (through a 20 year age gap) 
without being too old, as TikTok's adoption among those 55+ 
is low.
All accounts require age input. For all accounts, the birth dates were [Month] [Day], 2000 and [Month] [Day], 1980, where month and day were selected at random.

\subsubsection{Gender} 
Prior work on targeted advertising has shown that gender targeting can substantially increase ad engagement. For example, gender and interest targeted ads on Tumblr's achieved a 20\% lift in engagement compared to untargeted campaigns~\cite{grbovic2015gender}.
However this type of targeting could result in users of some genders being targeted disproportionately with beauty or weight loss advertisements which could negatively impact their mental health~\cite{blaine2002selling}. 

While TikTok's account creation does not enable direct gender input, users are prompted to share their name. Names are correlated with gender and many systems for inferring gender from names exist~\cite{10.1145/3568813.3600121}, so we tested two names (one man, one woman) for each account. 
Names were selected based on Social Security Administration data associated with the years selected: 
Jacob/Emily for the accounts aged 26, and  Michael/Jennifer for the 46 year old accounts~\cite{ssa_popular_names}. 


\subsection{Dependent Variables}
\label{section:method:dependent}

Frequency and content (e.g., advertisement category and price) of ads are our dependent variables. Screenshots of all content were captured and analyzed. 

\subsubsection{System Outputs} 
Screenshots were recorded for all content viewed by every account. While recording video IDs or URLs would provide the most data for analysis, the only way to access this information on TikTok is using the ``share'' function. 
Since sharing was one of our independent variables in Study 1, 
could be interpreted as signaling interest, and is unlikely to reflect typical user behavior when browsing, we used screenshots instead. Prior work has been able to obtain video IDs, but only through intercepting network traffic~\cite{10.1145/3589335.3651260}. Screenshots were taken immediately before scrolling to the next video. 

Recently in an effort to increase transparency, TikTok began offering an ``About This Ad'' explanation for each ad
~\cite{tiktok2022aboutthisad}. While the ad explanations have not been studied directly, prior work found that TikTok's explanations for which videos are shown were incorrect up to 40\% of the time
~\cite{mousavi2024auditing}. Due to this inaccuracy, as well as the fact that the menu to access the explanations is found in the share tab (which similarly risks contaminating our data collection), we do not collect or analyze these explanations. 


Each account was active for 3-4 days, or until at least 1,000 screenshots had been collected. Collection was stopped when all accounts running simultaneously reached 1,000+ consumed content. While users report having their algorithm return personalized results within minutes of joining the platform~\cite{10.1145/3589334.3645600}, we set the lower-limit of 1,000 consumed videos to increase the likelihood that the algorithm was personalized for the accounts. 

\subsubsection{Ad Detection} A two-step process was used to identify ads, starting with image processing and then, if the first step was unsuccessful, using OCR and keyword matching. This was due to the variability of visual representations and keywords of TikTok ads. 
For each image, the presence of a red call-to-action button is first detected. If no button is found, OCR-extracted keywords such as \texttt{sponsored}, \texttt{creator earns commission}, and \texttt{ad} are used to classify the content. The method achieved 99.6\% accuracy on a balanced sample and 100\% accuracy on a random sample of screenshots. Additional details and the full keyword list are provided in Appendix B. Of the 80,169 screenshots, 24,255 (30.2\%) were classified as ads. 

\subsubsection{Advertisement Content Categorization} 
Automatically detected logos and image labels (described below) were used to categorize the content of each ad. 
Due to our focus on advertising and commerce, all 49 final categories were generated on the basis of the logos. These included categories like \texttt{Restaurants \& Fast Food Chains}, \texttt{Beauty \& Cosmetics}, \texttt{Fashion \&
Apparel}, and \texttt{Sports}. The 445 unique logos were first grouped into 52 higher-level categories using large language model (LLM) prompting, with manual oversight (i.e., correcting, removing, and merging) over the final categories. 
For example, the \texttt{Restaurants \& Fast Food} category included logos for brands including \texttt{Sweetgreen}, \texttt{Jersey Mike's Subs}, and \texttt{IHOP}. 
Appendix C 
provides the full prompt and 
details of the 
annotation process. 
A second LLM prompt 
mapped the 2,866 image-based labels to the same 49 categories. 563 labels (e.g., people, happiness, beard) were classified as ``Other'' due to their ambiguity. 

Each ad was 
assigned a single category. 
If a logo was present, the ad was assigned the most common category among its logos (2,396 ads). 
If no logos were detected, 
the ad was assigned the most common category among the image labels (8,906 ads). 
If neither logos nor image labels were identified, the image was 
classified as ``Unknown'' (345 ads). 

\vspace{3pt}
\noindent \textit{Logo Detection} Logos were detected using 
\textit{Google Cloud Vision API} logo detection on all 
advertisements and interacted-with content. Logos allow us to identify brands and industries that advertised on TikTok. 
Of the 11,647 ads in Study 2, 2,396 
(20.6\%) 
contained at least one detected logo. 
This processing identified 445 unique logos. 

\vspace{3pt}
\noindent \textit{Image Label Detection} Google Cloud Vision API's image label function identified the visuals in ads, providing signals about the advertisement's industry when obvious information (e.g., logo) was unavailable. All images were cropped to avoid everything being labeled a screenshot, however, we still perform post processing of labels to remove tags such as \texttt{screenshot} and \texttt{advertisement} when present. Appendix D includes additional details.

\vspace{1pt}
\subsubsection{Price Detection} Users of different income groups could be targeted with differently priced items and advertisements. 
In Study 2, to understand how the prices of ads differ by income, all ads that contained \$+Number or Number+\% were extracted (2,096 ads, 18\%). One member of the research team manually reviewed and labeled these ads, distinguishing instances where  dollar amounts were discounts or deals rather than product prices. A similar separation, splitting discounts from rates (e.g., a lender's APR) was done for percentages. Incorrect detections, as well as cases in which the \$/\% referred to a hypothetical income or use of a product (e.g., ``if you build a website with us, you could make \$X''), were also removed, leaving 1,675 ads. 
While some discounts (e.g., ``Buy 1 Get 1 Free'') are missed, this approach captured any ad with explicit prices (i.e., any price preceded by a \$ sign) or discounts. 



\subsection{Summary of Experimental Conditions} 

\subsubsection{Study 1.} To explore the effects of interaction type we created 32 accounts: 1 device type $\times$ 2 ages $\times$ 2 names $\times$ 4 interaction types $\times$ 2 copies. 

\subsubsection{Study 2.} To investigate the effects of device price we created 24 accounts: 3 device types $\times$ 2 ages $\times$ 2 names $\times$ 1 interaction type $\times$ 2 copies. 

While data associated with gender and age were collected in both studies, due to the artificial interaction behaviors in Study 1, the effects of gender and age were explored only using Study 2 data. 
Data on accounts within one setting (e.g., ``Jacob'' accounts performing different interaction types) were collected simultaneously to control for any 
marketing campaigns. 
Two accounts (i.e., two copies) were created for each setting to control for randomness in recommendations and targeting. 

\subsection{Auditing System Implementation} 

\subsubsection{Software} We developed a python and ADB script to automate interactions with the TikTok application on Android devices. Code will be provided to researchers upon request.

\subsubsection{Hardware} We selected identical low, medium, and high price devices (four of each kind) to run our automated accounts. The different prices allowed us to explore the effects of price on the advertisements received. Having four of each device allowed us to conduct data collection in parallel, helping us capture time-restricted advertising campaigns as consistently as possible across all experimental conditions.

Android devices were used because they offer greater support for large-scale automation and scripted interactions. Apple devices impose restrictions making automated auditing and resetting of ad IDs difficult to implement.


\subsubsection{Switching Between Accounts} After logging out of each account and prior to creating the new one, we deleted the TikTok app storage and reset the device's advertiser ID. These steps were taken to reduce the likelihood that the platform could link the accounts with each other and recommend similar content or advertisements on them. Account creation, login, and logout were all performed manually due to TikTok's bot-detection mechanisms, which made automating these steps unreliable. All other interactions with the TikTok feed were performed automatically. 

\subsection{Data Availability}
\label{section:data-availability}
Across both studies, we have collected 24,255 advertisement screenshots. Researchers can request access to the data by filling out a form\footnote{\url{https://forms.gle/6HEVqeMJBqotyrWd8}} 
or contacting the first author. 
The dataset will not be made publicly available online because in addition to corporate advertisements, some users place ads to increase the reach of their individual user accounts, 
which often include identifiable information such as usernames. 





\section{RQ1: How does interaction type affect the number and content of ads?}

\subsection{Liking and sharing increased ad load} 

\begin{table*}
    \centering
    \begin{tabular}{l l l  r r  r r  r r  r r}
        \textbf{Age} & \textbf{Name} & \textbf{} & \multicolumn{2}{c}{\textbf{Scroll}} & \multicolumn{2}{c}{\textbf{Like}} & \multicolumn{2}{c}{\textbf{Comment}} & \multicolumn{2}{c}{\textbf{Share}}\\\hline 
        \multirow{2}{*}{26}  &  \multirow{2}{*}{Jacob}  &  1  & 1400 & \textit{26.0\%} & 1016 & \textit{30.6\%} & 1451 & \textit{27.6\%} & 1347 & \textit{32.8\%}\\
            &         &  2  & 1196 & \textit{20.2\%} & 1324 & \textit{36.6\%} & 1666 & \textit{20.7\%} & 1280 & \textit{34.9\%}\\[0.5em]
        \multirow{2}{*}{26}  &  \multirow{2}{*}{Emily}  &  1  & 1065 & \textit{25.0\%} & 1509 & \textit{31.1\%} & 1402 & \textit{25.4\%} & 1038 & \textit{33.0\%}\\
            &         &  2  & 1865 & \textit{25.4\%} & 1372 & \textit{31.3\%} & 1312 & \textit{23.8\%} & 2127 & \textit{37.6\%}\\[0.5em]
        \multirow{2}{*}{46}  &  \multirow{2}{*}{Michael}  &  1  & 1341 & \textit{26.6\%} & 1007 & \textit{29.9\%} & 1222 & \textit{26.9\%} & 1062 & \textit{35.8\%}\\
            &           &  2  & 1621 & \textit{27.7\%} & 1184 & \textit{24.5\%} & 1472 & \textit{30.5\%} & 1598 & \textit{36.3\%}\\[0.5em]
        \multirow{2}{*}{46}  &  \multirow{2}{*}{Jennifer}  &  1  & 1487 & \textit{26.9\%} & 1250 & \textit{30.1\%} & 1025 & \textit{26.0\%} & 1460 & \textit{35.5\%}\\
            &            &  2  & 1220 & \textit{27.6\%} & 1230 & \textit{31.9\%} & 1024 & \textit{26.2\%} & 1246 & \textit{36.0\%}\\[0.5em]
        \hline
        \rowcolor{pink!20} \multicolumn{3}{r}{Mean} & 1399 & \textit{25.7\%} & 1236 & \textit{30.8\%} & 1322 & \textit{25.9\%} & 1395 & \textit{35.2\%}\\
        \rowcolor{pink!20} \multicolumn{3}{r}{Median} & 1370 & \textit{26.3\%} & 1240 & \textit{30.9\%} & 1357 & \textit{26.1\%} & 1314 & \textit{35.6\%}\\
        \rowcolor{pink!20} \multicolumn{3}{r}{Standard Deviation}& 257 & \textit{2.4\%} & 171 & \textit{3.3\%} & 224 & \textit{2.8\%} & 350 & \textit{1.6\%}\\
    \end{tabular}
    \caption{Study 1 result. Total number of videos captured and ad load per account \& interaction type. Liking and sharing lead to the highest ad load. Surprisingly, commenting which is a very active form of engagement, did not result in increased ad load.}
    \label{tab:stats_of_study_1_ads}
\end{table*}

Ad load varied considerably for the different interaction types (like, comment, share, scroll). Table \ref{tab:stats_of_study_1_ads} displays the number of videos viewed 
and ad load for 
each account, with two unique accounts created for each demographic attribute. 
Surprisingly, \textit{commenting}, which is a very active form of engagement, 
did not increase exposure to ads compared to the scroll condition. \textit{Like} and \textit{Share} conditions attracted the highest proportions of ads, with overall means of 30.8\% and 35.2\% respectively, compared to 25.7\% for \textit{scroll} and 25.9\% for \textit{comment}. The \textit{share} condition produced the highest ad rate across all participants and rounds, with individual session rates ranging from 32.8\% to 37.6\%. Average ad load of 29.4\% across all interaction conditions is comparable to cable at 13.3 minutes per hour~\cite{Friedman2021TVAdLoadSpike}, but more than most streaming services~\cite{Sutton2022StreamingAdLoads}.


\begin{table}
\centering
\begin{tabular}{lrrl}
Predictor & $\beta$ Coefficient & $p$-value & \\
\hline
$\beta_0$ Intercept & -113.0755  & 0.0141  &$^{*}$\\
$\beta_1$ \# Videos Watched & 0.3384 &  0.0000 & $^{***}$\\
$\beta_2$ Liked  &  75.9950 & 0.0009 & $^{***}$\\
$\beta_3$ Shared  & 135.1902 & 0.0000  & $^{***}$ \\
$\beta_4$ Commented  & 5.7698 & 0.7756 & \\
\hline 
\multicolumn{4}{l}{%
  $n = 32$, $R^2 = 0.8847$, Adj.\ $R^2 = 0.8676$} \\
\multicolumn{4}{l}{\footnotesize Significance: $^{***}p<0.001$\ \ $^{*}p<0.05$} \\
\end{tabular}
\caption{OLS regression model predicting the number of ads based on interaction and demographic variables. Liking and sharing significantly increased ad count, while commenting had a positive but non-significant relationship.}
\label{tab:ols_single}
\end{table}

To test for statistical significance of the differences observed between interaction types we fit an OLS regression model on the data extracted in Study 1: $\widehat{\texttt{\#Ads}}_i =
\beta_0
+ \beta_1 \cdot \texttt{\#Videos Watched}_i
+ \beta_2 \cdot \texttt{Liked}_i
+ \beta_3 \cdot \texttt{Shared}_i
+ \beta_4 \cdot \texttt{Commented}_i
$, for account $i \in \{1,...,32\}$ accounts. where \texttt{\#Videos Watched} is the total number of videos consumed on the account. Interaction type is represented by three indicator variables — \texttt{Liked}, \texttt{Commented}, and \texttt{Shared} — each coded 1 if the account performed that interaction and 0 otherwise. \texttt{Scroll} serves as the reference category, so when all three interaction indicators are 0, the model reflects the scroll condition.


Liking and sharing significantly increased ads. Commenting exhibited a small positive, but non-significant, relationship with ad count. Overall, the model confirmed the significance of interaction type as a predictor of ad exposure, even after controlling for demographic variables and session volume. Table \ref{tab:ols_single} shows the results of the model. The model explains approximately 88.5\% of variance in ad counts. The findings persisted in a model that included age and gender (Appendix F).

\subsection{As accounts viewed more content, they were shown more advertisements}

\begin{figure}
    \centering
    \includegraphics[width=\linewidth]{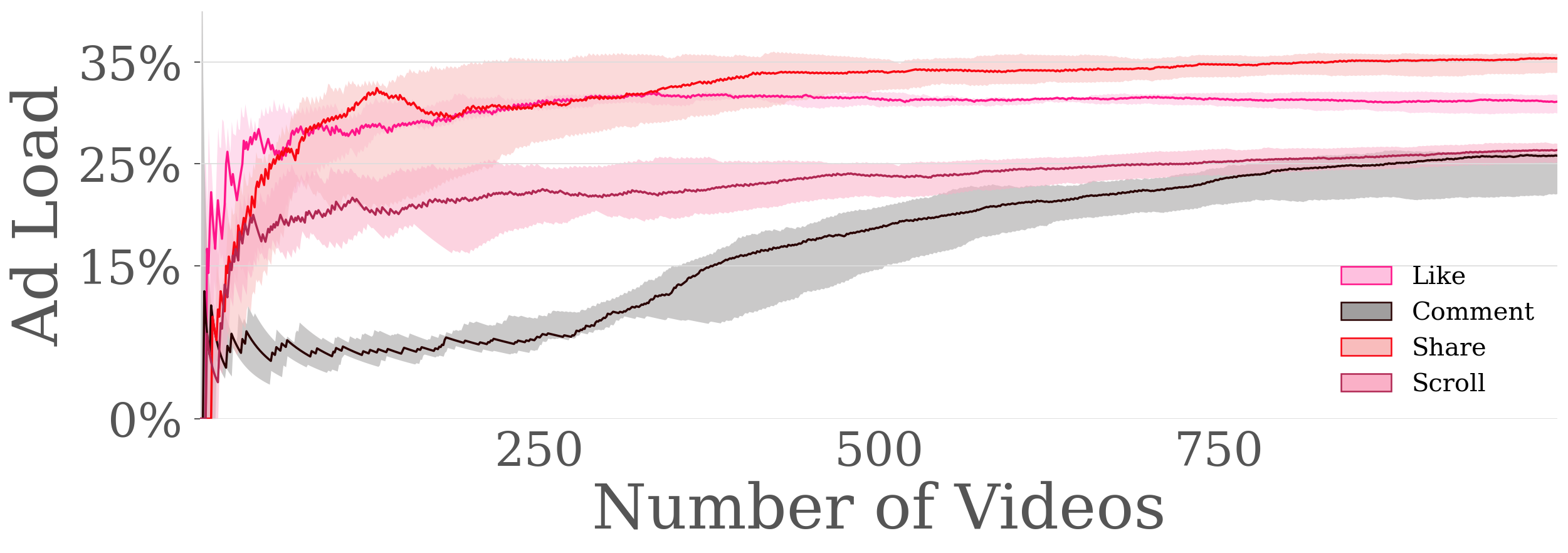}
    \caption{Ad rate over time. All accounts experienced an increased ad load as more content was viewed. The comment accounts were unusual in that the increase was delayed relative to other interaction types.
    }
    \label{fig:number_of_ads_by_interval}
\end{figure}

As accounts encountered more content, the ad load increased. Figure \ref{fig:number_of_ads_by_interval} shows this increase over the lifetime of each account. For each account, 
the proportion of ads in the first 1,000~videos was calculated for each step$_\textit{n}$ 
by dividing the cumulative number of ads observed up to video$_\textit{n}$ by \textit{n} (the 
number of videos viewed). For consistency, any videos after the first 1,000 were discarded in this analysis. 
Median and first and third quartiles are shown in Figure \ref{fig:number_of_ads_by_interval}.

A general upward trend for all interaction types was observed. 
This result indicates that TikTok's ad delivery is not uniformly distributed across an accounts' lifetime. Instead, the platform appears to serve advertisements at an increasing rate as consumption continues, independent of how the user is engaging with content. One interpretation is that the algorithm treats sustained scrolling itself as a signal of engagement, and increases ad load accordingly. Alternatively, the pattern may reflect a deliberate pacing strategy in which initial content is kept relatively ad-light to establish momentum before monetizing the session more aggressively. 

Interestingly, commenting exhibited a different pattern from other interactions. While there was an eventual increase in ad load, it occurred later and at a much slower rate. This may be due to only commenting being categorized as an unusual behavior by the platform. We leave further investigation of how different interaction start times, interaction behaviors, or interaction rates affect ad load to future work.

\subsection{More interactions with ads increased ad load}

\begin{figure}
    \centering
    \includegraphics[width=0.9\linewidth]{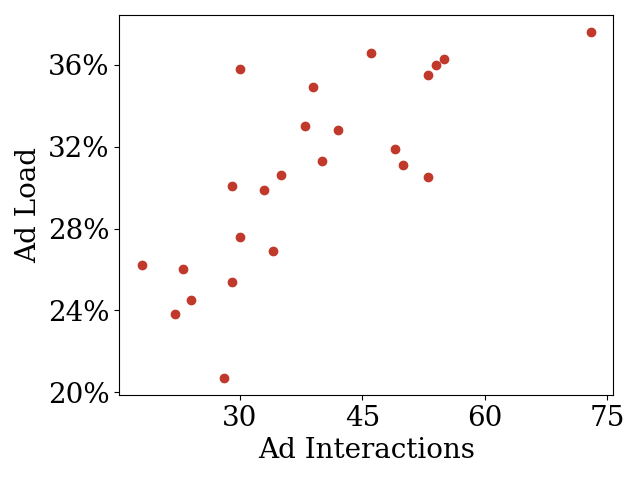}
    \caption{The relationship between ad load and the number of interactions with ads across Study 1 accounts. As the number of interactions with ads increases, so does the percentages of advertisements encountered.} 
    \label{fig:rq1_paired_figure}
\end{figure}

Users' interaction with specific types of content could explain the prevalence of content they see. In other words, accounts that interact with more ads over their lifetime, might be exposed to more ads. 
Both ads and non-ads were interacted with across all accounts. Figure \ref{fig:rq1_paired_figure} visualizes the relationship between ad load and number of ads interacted with. 

When accounts interacted with more ads, their ad load increased. The Pearson correlation between the values displayed is +0.759 ($p < 0.001$). Thus, both the type (Table \ref{tab:ols_single}) and target of the interaction impacted ad load.


%


\section{RQ2: How does income affect the number and content of ads?}

\subsection{More expensive devices received fewer ads}

In our second study, realistic user interactions were simulated to investigate the impact of income, approximated through device price, on ad load and content. The mean and median total number of content consumed and ad load for all accounts on each device\footnote{A total of 8 unique accounts are run on each device: 2 ages $\times$ 2 genders $\times$ 2 copies.} are displayed in Table \ref{tab:stats_per_device}.

\begin{table}
    \centering
    \begin{tabular}{l r r r r}
        \textbf{Device} & \multicolumn{2}{c}{\textbf{Mean}} & \multicolumn{2}{c}{\textbf{Median}}\\\hline
        Pixel 3a & 1493 & \textit{33.9\%} & 1514 & \textit{34.1\%}\\
        Pixel 8 & 1531 & \textit{31.5\%} & 1442 & \textit{32.1\%}\\
        Galaxy S26+ & 1644 & \textit{29.1\%} & 1624 & \textit{30.4\%}\\
    \end{tabular}
    \caption{Device price \& ad load. More expensive devices had slightly lower ad loads.}
    \label{tab:stats_per_device}
\end{table}




Our results indicated that the least expensive device received the highest ad load (
33.9\%), followed by the mid-priced device (
31.5\%), and the most expensive device had the lowest ad load (
29.1\%). These findings provide support for the hypothesis that device price influences ad exposure.


To further investigate the factors driving ad exposure, we fit an OLS regression model predicting the number of ads received: $\widehat{\texttt{\#Ads}}_i =
\beta_0
+ \beta_1 \cdot \texttt{\#Videos Watched}_i
+ \beta_2 \cdot \texttt{Age}_i
+ \beta_3 \cdot \texttt{Gender}_i
+ \beta_3 \cdot \texttt{Device Price}_i
$, for account $i \in \{1,...,24\}$ accounts. The reference group for \texttt{Age} is 26 and for \texttt{Gender} is Man. Device price is coded in U.S. dollars. Table \ref{tab:regression_results_continuous_price} displays the results of this model.

Device price had a negative, but non-significant relationship with ad count. Age was the only significant demographic variable, with younger profiles receiving approximately 89 more ads per account than older profiles (a 5\% increase in ads).

\begin{table}[ht]
\centering
\begin{tabular}{l r r r}
\textbf{Predictor} & \textbf{$\beta$ Coefficient} & \textbf{$p$-value} \\
\hline
$\beta_0$ Intercept & -41.4320 &  0.7789 \\
$\beta_1$ \# Videos Watched & 0.3263 & 0.0018 & $^{**}$ \\
$\beta_2$ Age & 89.1423 & 0.0329 & $^{*}$\\
$\beta_3$ Gender & 43.2788 & 0.2720 \\
$\beta_4$ Device Price & -0.0943 & 0.1149 \\
\hline
\multicolumn{4}{l}{%
  $n = 24$, $R^2 = 0.4608$, Adj.\ $R^2 = 0.3473$} \\
\multicolumn{4}{l}{\footnotesize Significance: $^{**}p<0.01$\ \ $^{*}p<0.05$} \\
\end{tabular}
\caption{OLS regression model predicting the number of ads based on device and demographic variables. Age is the only statistically significant demographic variable, with younger users receiving more ads. Device price had a negative but not significant relationship with number of ads.}
\label{tab:regression_results_continuous_price}
\end{table}

\subsection{Travel ads targeted wealthy consumers, but most other categories were based on non-ad interactions} 

\begin{figure}[t]
    \centering
    \includegraphics[width=0.97\linewidth]{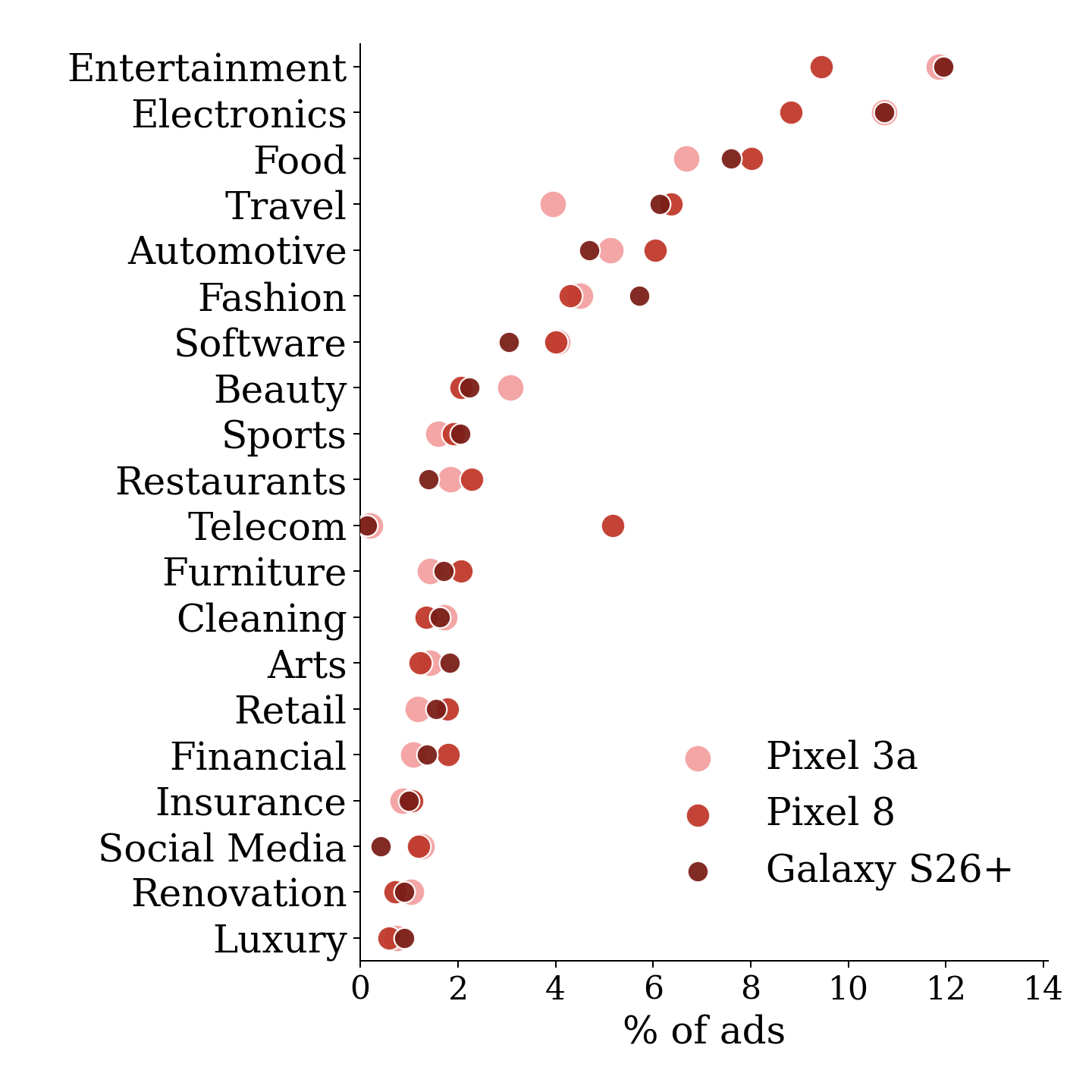}
    \caption{Category distribution of advertisements across the three phone types. Entertainment and Electronics were the most common categories across devices. Travel ads appeared less frequently on the cheapest device (Pixel 3a) than on the more expensive phones.}
    \label{fig:phones_cropped_top_labels_ads}
\end{figure}

Entertainment, electronics, and food were among the most frequently advertised categories (Figure \ref{fig:phones_cropped_top_labels_ads}), 
across all device types. While many of these categories are consistent with prior work, TikTok seems to feature significantly less healthcare advertising than other media
~\cite{marketingcharts_top_ad_categories_2020}. 
Alternatively, the relevant visual features of categories such as healthcare may have been 
harder to capture or unavailable in our screenshots. Because healthcare ads often feature people, and our ad categorization method removed visual categories such as people and happiness due to their ambiguity, some advertisements in this category may be missed.

Travel was among very few categories which received noticeably more ads in 
the more expensive phones. Otherwise most categories were similarly frequent for accounts on all phones, and pairwise Wilcoxon signed-rank tests conducted on the ranked lists of ad categories across pairs of devices found no statistically significant differences. 




\begin{figure}
    \centering
    \includegraphics[width=0.965\linewidth]{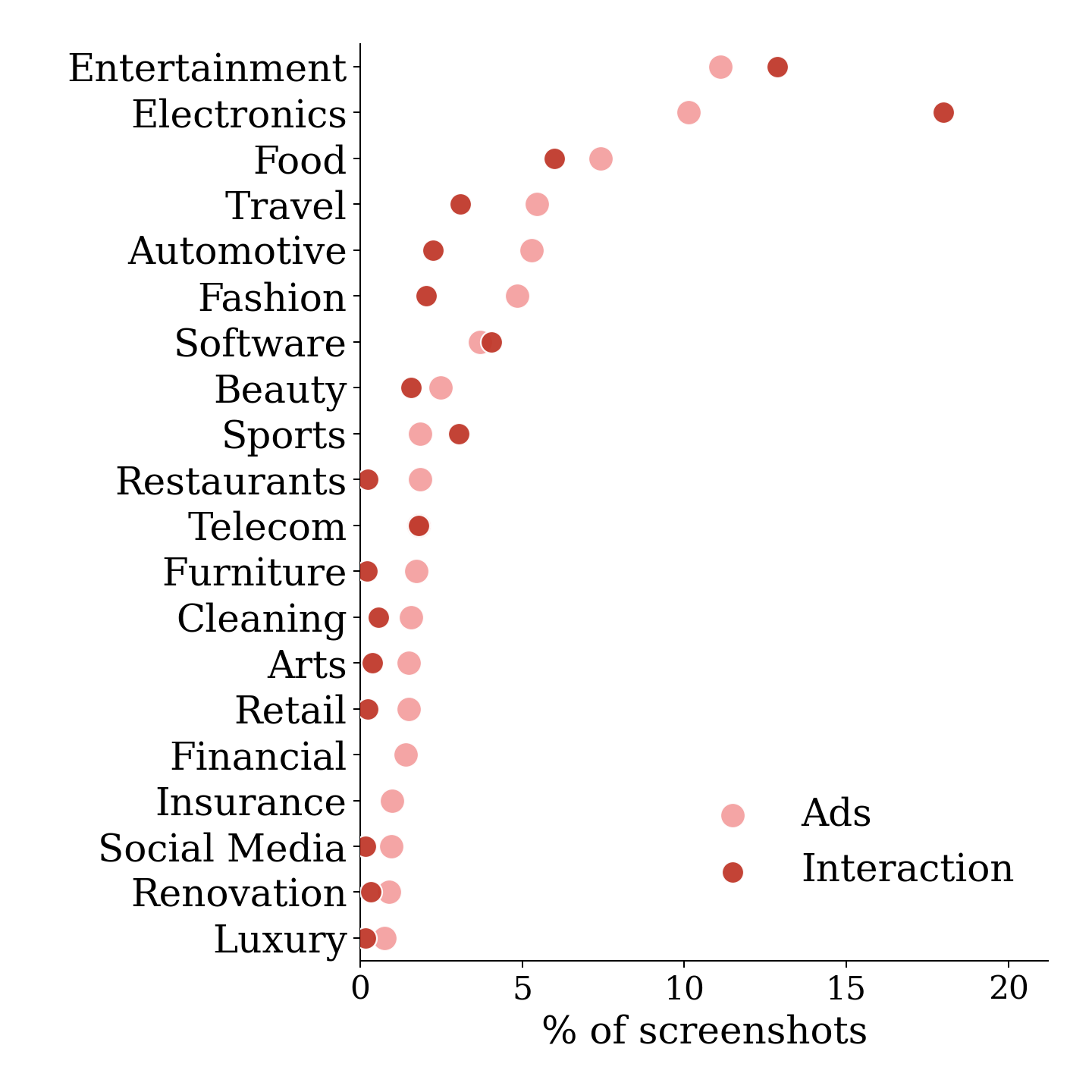}
    \caption{Category distribution of advertisements across advertisements and interacted-with content. For many categories, increased interaction led to an increase in ads of that type. However, accounts received advertisements for some categories (e.g., Restaurants, Furniture, Arts) that had little to no interactions. 
    }
    \label{fig:interacted_vs_seen}
\end{figure}

Instead, rather than device price, accounts' interactions with non-ad content was most predictive of the ad categories. For many categories, increased interaction led to an increase in ads of that type. However, accounts received advertisements for some categories (e.g., Restaurants, Furniture, Arts) that had little to no interactions (Figure~\ref{fig:interacted_vs_seen}). 
Pairwise Wilcoxon signed-rank tests 
on the 
ad and interaction categories 
were statistically significant (W=299.00, p$<0.001$). Thus, it appears that some business sectors place ads to uninterested users; future study is needed to test this. 


\subsection{The cheapest phone was shown the highest discount percentages}

Advertising the exact price of goods was relatively uncommon across devices (14.4\% of total ads in Study 2). The most expensive phone received the fewest discounts in ads. Table \ref{tab:price_differences_between_phones} presents the median cost and discount of ads per device.


         

\begin{table}[!ht]
    \centering
    \resizebox{\linewidth}{!}{
    \begin{tabular}{r r r r r r r}
    
    & \multicolumn{2}{c}{\textbf{Pixel 3a}} & \multicolumn{2}{c}{\textbf{Pixel 8}} & \multicolumn{2}{c}{\textbf{Galaxy S26+}}\\\hline

    \textbf{Cost} &  236 & \$31.20 & 369 & \$53.30 & 337 & \$50.37\\
    \textbf{Discount} & 245 & 45.00\% & 296 & 28.50\% & 165 & 20.00\%\\
    \textbf{Total Ads} & 4,059& & 3,771&  & 3,817 &\\
    

         
    \end{tabular}
    }
    \caption{Median costs and discounts of ads across devices. Mid-range device received the most expensive ads. The cheapest device received the highest discounts.}
    \label{tab:price_differences_between_phones}
\end{table}

Accounts on the cheapest phone were shown the highest discount percentages. It is important to note that this does not necessarily translate to cheaper deals (many ads featuring discounts do not include the product cost), only that the advertisement presented itself as offering a greater discount to the user. Accounts on more expensive devices were shown advertisements for higher cost goods and services, consistent with an assumption of higher spending power by users on these devices. Surprisingly, however, higher price advertisements were shown on the mid-range device compared to the most expensive device.


\section{RQ3: How do age and gender affect the number and content of ads?}

\subsection{Younger accounts received more ads}

To assess the impacts of age and gender on ad load we compared mean and median ad loads across accounts of the different age and gender in Study 2, shown in Table \ref{tab:stats_per_demographic}. We found that younger users and men received more ads. However, age was the only statistically significant predictor of ad count between the two (Table \ref{tab:regression_results_continuous_price}). This finding was consistent with prior work which found a difference in content recommendation across age, but found no significant difference between genders~\cite{10.1145/3589335.3651260}. 

Given the low quality of name-based gender detection algorithms, specifically for racial minorities and non-binary users~\cite{10.1145/3568813.3600121}, the apparent lack of platform reliance on this feature for ad targeting is a positive finding. However, the lack of observed gender differences may be due to TikTok choosing not to use names as a proxy for gender, given the low accuracy of such inferences and the ease of providing false names. TikTok may still rely on other signals to infer gender, which future work should investigate.




\subsection{Ad categories were consistent across age and gender}

To assess the differences in the content of advertisements, we examined the differences in the prevalence of advertisement category rankings between different ages and genders. Pairwise Wilcoxon signed-rank tests conducted on the ranked lists of ad categories found no statistically significant differences for either factor.

\begin{table}
    \centering
    \begin{tabular}{l r r r r}
         & \multicolumn{2}{c}{\textbf{Mean}} & \multicolumn{2}{c}{\textbf{Median}}\\\hline
         \textit{Gender} & & \\\hline 
        Man & 1521 & \textit{32.9\%} & 1562 & \textit{32.1\%}\\
        Woman & 1592 & \textit{30.1\%} & 1674 & \textit{30.8\%}\\\hline 

        \textit{Age} & & \\\hline 
        26 & 1507 & \textit{34.5\%} & 1562 & \textit{35.0\%}\\
        46 & 1605 & \textit{28.4\%} & 1648 & \textit{30.4\%}\\
    \end{tabular}
    \caption{Demographics \& ad load. Men had slightly higher ad load. Younger accounts had higher ad load.}
    \label{tab:stats_per_demographic}
\end{table}

\section{Discussion}
\label{section:discussion}

\subsection{A focus on behavior rather than demographics}

Our finding revealed that active engagement (i.e., liking and sharing) substantially increases the rate at which users were served advertisements. This pattern is consistent with prior work on content recommendation systems, where engagement signals have been found to influence content personalization~\cite{10.1145/3485447.3512102}. On one hand, this type of engagement-based targeting provides users with more power to influence the ads they get by strategically interacting with content, on the other hand it creates an adversarial environment in which users refrain from supporting content or creators they enjoy out of a fear of receiving excessively high volumes of similar content~\cite{10.1145/2858036.2858494}.


Given the importance of behavioral factors future studies of the topic are also required. In this work, interactions were performed at a constant rate of 10\%, yet we found that the percentage of ads presented to users increased over time. One interpretation is that TikTok's ad delivery algorithm treats longer engagement with the application as a signal or investment or buy-in, and increases ad load. An alternative explanation could be a feedback-loop dynamic: as accounts were shown more ads, the 10\% random interaction probability meant they were statistically more likely to also interact with ads, further reinforcing ad delivery. Future work should investigate different engagement patterns with ads and non-ad content. Additionally, impacts of changing behavior over time should also be examined.

\subsection{Device price as a weak and ambiguous signal}

Our second study found that the cheapest device had the highest ad load. This means that users who do not have the means or desire to purchase more expensive devices may face a higher advertisement burden, with ads taking up more of their time and attention. While differences in ad load and content were small, given the unfairness of such differences, we believe further investigation is required. Thus, future work should continue examining features that may act as proxies for income and how platforms use them. For example, our work explored TikTok's Device Price feature, but other features such as Spending Power remain unexplored.

\section{Conclusion}

This paper presented two paired studies 
auditing TikTok's advertising system, examining the effects of interaction type, device price (as an income proxy), age, and gender on the number and content of ads. 
Across the two studies, using 56 accounts, we collected data on 80,169 videos 
and 24,255 advertisements. We found that active engagement, specifically liking and sharing content, significantly increased ad exposure. Commenting, however, did not increase ad load. Lower-income accounts were shown higher discount percentages. Device price was also found to have a directional effect with the least expensive device receiving more ads.



\bibliography{main}

\newpage



\appendix
\section{TikTok Advertisement Platform}
\label{appendix:section:tikotk_stats}





TikTok offers an advertising platform\footnote{\url{https://ads.tiktok.com/}} which allows advertisers to select the characteristics of users (e.g., location, age, interests, among many others) they would like to target. The platform provides advertisers with an \texttt{Available audience} estimate which is an ``estimation of the number of user accounts eligible'' to see the ad based on the selected criteria. This estimate is ``the number of user accounts shown ads [on TikTok] in the past 30 days''. 

Among the demographics and variables on the website are \texttt{Age} and \texttt{Device Price}. The age groups on the platform are as follows: 13-17, 18-24, 25-34, 35-44, 45-54, and 55+. The website, however, does not provide estimates for user counts under the age of 18. The minimum and maximum value of device prices can be selected in \$50 increments. The highest value advertisers can select is 1000+. We collected audience estimates on \texttt{March 10, 2026}.

Table \ref{tab:user_estimates_for_tiktok_os} displays the gender distribution of TikTok users overall and those using Android devices. We can see that iOS users are more common. However, our automated interaction system only operated on Android devices.

Advertisers can target users based on income using two options: (i) Device price (Table \ref{tab:user_estimates_for_tiktok_age_and_price}) and (ii) Spending power (Table \ref{tab:user_estimates_for_tiktok_spending_power}). We focus on device price in this work. It is unclear to us what the criteria for ``spending power'' classification are. Table \ref{tab:user_estimates_for_tiktok_age_and_price} shows the number of users using devices of specific prices. We find that the values are different if the advertiser selects \texttt{any price} or \texttt{specific price} and selecting 0-1000+ as the range. This might be because TikTok is unable to determine the price of some user's devices. We selected the device price ranges shown in the table for our study because the number of users within the three groups are roughly similar.

\begin{table}[]
    \centering
    \begin{tabular}{c|c|c}
        \textbf{Gender} & \textbf{OS} & \textbf{Available audience}\\\hline 
        All & All & 146,115,000-178,586,000\\
        Male & All & 76,208,000-93,144,000\\
        Female & All & 69,907,000-85,443,000\\\hline
        All & Android & 36,889,000-45,087,000\\
        Male & Android & 22,981,000-28,089,000\\
        Female & Android & 13,907,000-16,998,000\\
    \end{tabular}
    \caption{Estimate of users located in the US ($Age \geq 18$). OS = operating system.}
    \label{tab:user_estimates_for_tiktok_os}
\end{table}

\begin{table}[]
    \centering
    \begin{tabular}{c|c}
        \textbf{Device Price Range} & \textbf{Available audience}\\\hline 
        Any &  36,889,000-45,087,000\\
         0-1000+ &  7,177,000-8,773,000\\\hline
        0-250 &  2,636,000-3,223,000\\
        400-650 &  899,000-1,100,000\\
         750-1000+ &  1,845,000-2,256,000\\
        
    \end{tabular}
    \caption{Estimate of users located in the US and using Android devices of various prices ($Gender = All$, $Age \geq 18$).}
    \label{tab:user_estimates_for_tiktok_age_and_price}
\end{table}

\begin{table}[]
    \centering
    \begin{tabular}{c|c|c}
        \textbf{SP} & \textbf{OS} & \textbf{Available audience}\\\hline 
         All & All & 146,115,000-178,586,000\\
        All & iOS & 109,226,000-133,499,000\\
         All & Android & 36,889,000-45,087,000\\\hline 
         High & All & 24,638,000-30,114,000\\
        High & iOS & 18,159,000-22,195,000\\
        High & Android & 6,478,000-7,919,000\\
    \end{tabular}
    \caption{Estimate of users located in the US and any gender and any device price ($Age \geq 18$). SP = Spending power, OS = operating system.}
    \label{tab:user_estimates_for_tiktok_spending_power}
\end{table}

\section{Ad Detection}

We perform a two-step ad identification process. In this section, we will describe what ads can look like, our identification process, and the performance of our method. 

\subsection{What do ads look like?}

\begin{figure*}
    \centering
    \begin{subfigure}[b]{0.48\textwidth}
    \includegraphics[width=0.31\linewidth]{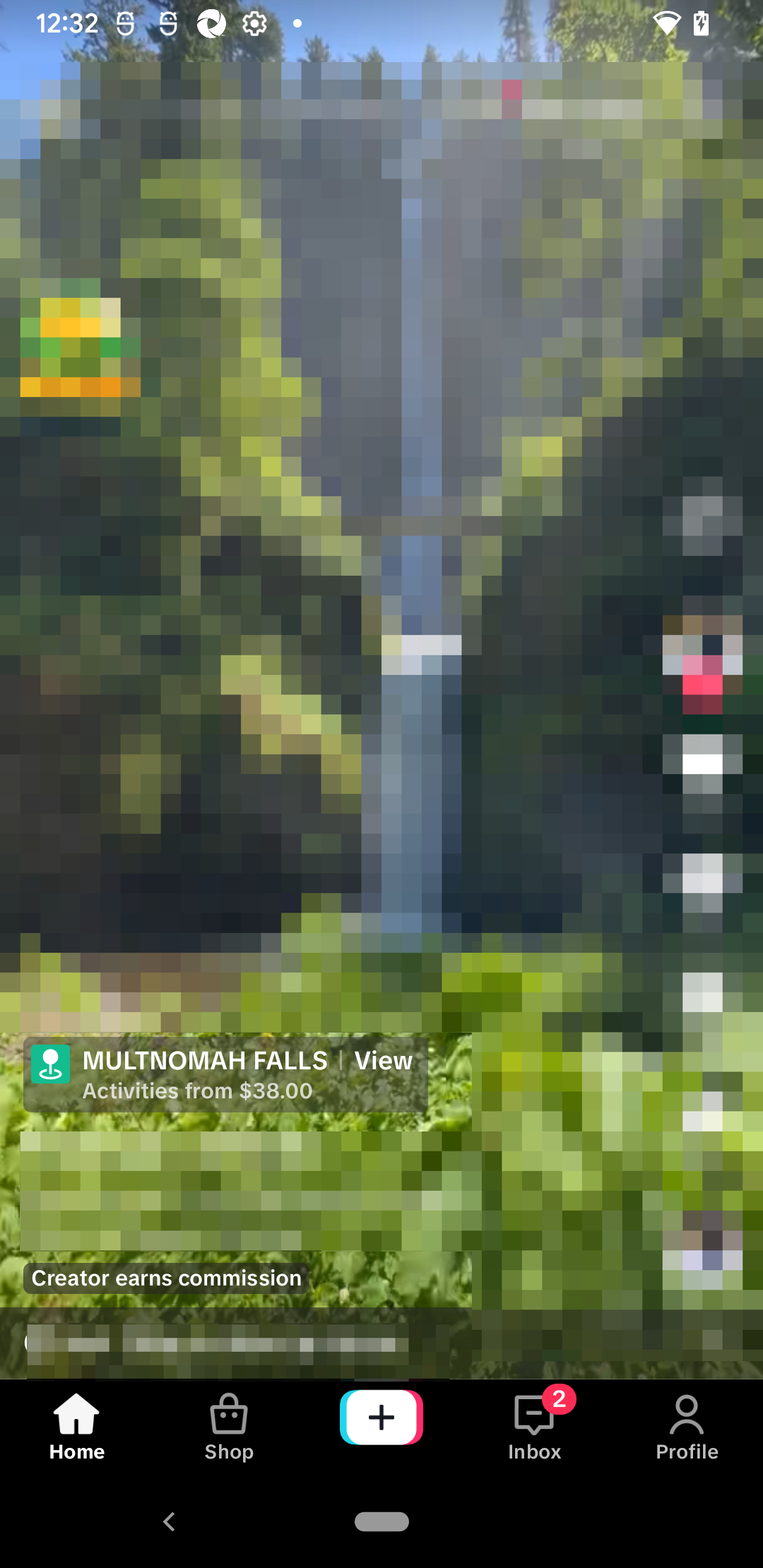}
    \includegraphics[width=0.3\linewidth]{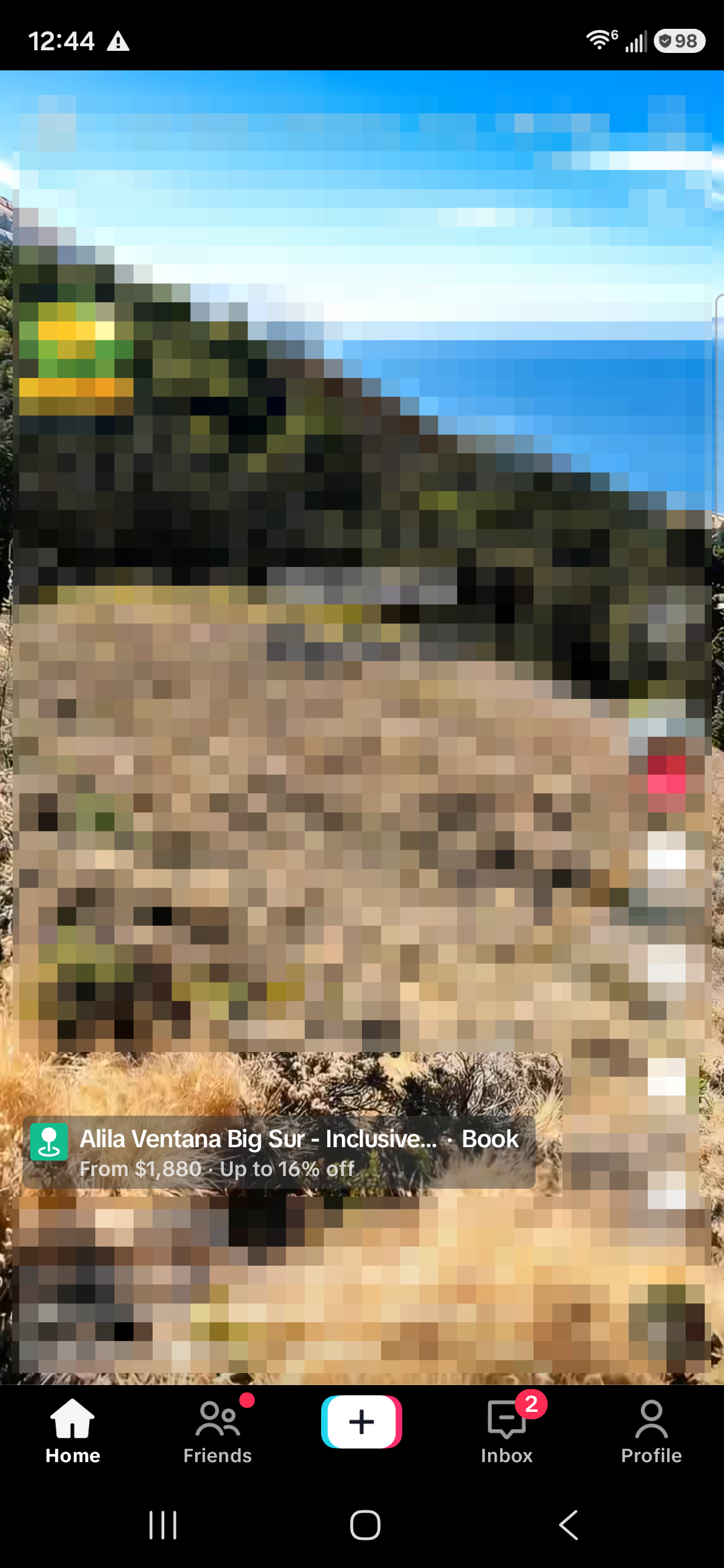}
    \includegraphics[width=0.3\linewidth]{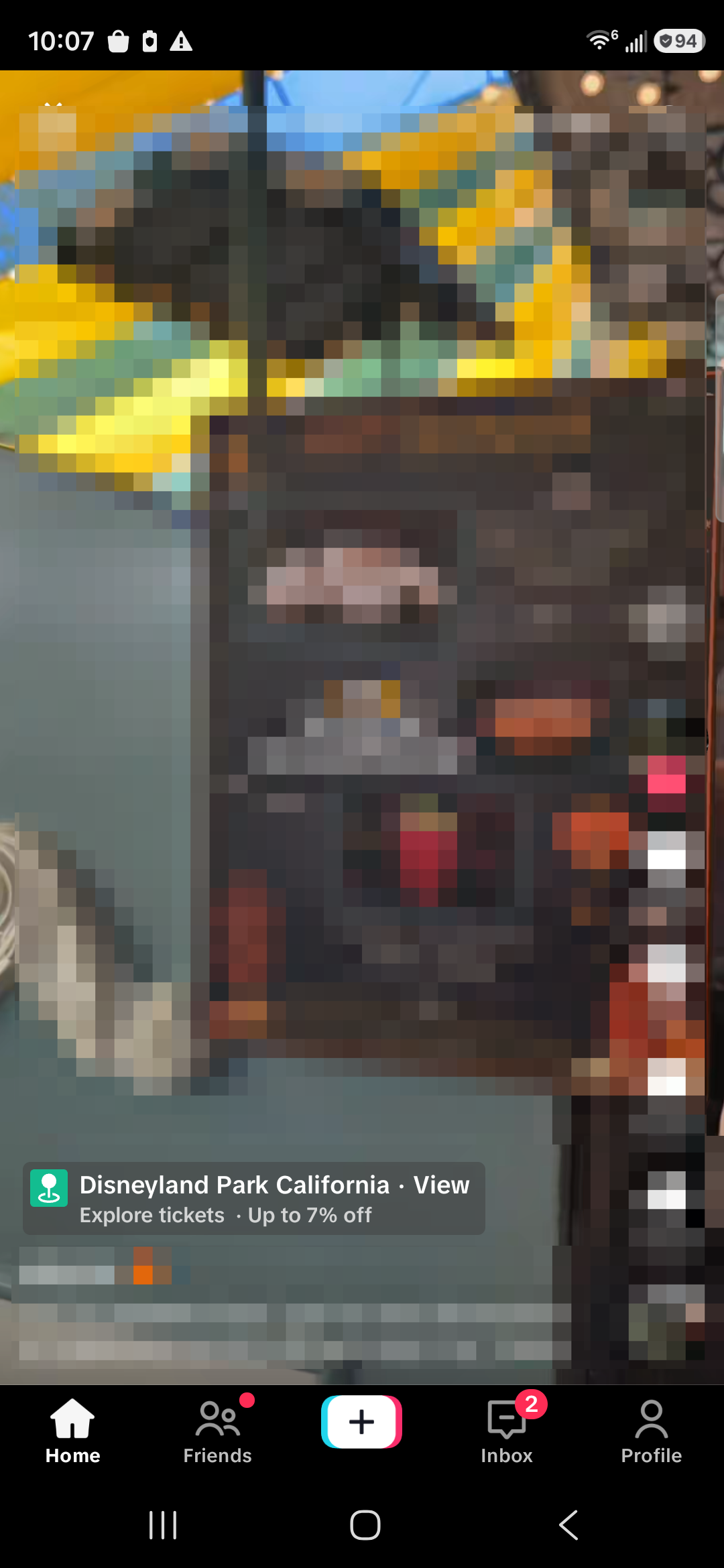}
    \caption{Multiple tags might be present.}
        \label{fig:ad_multiple_tags}
    \end{subfigure}
    \begin{subfigure}[b]{0.48\textwidth}
    \includegraphics[width=0.3\linewidth]{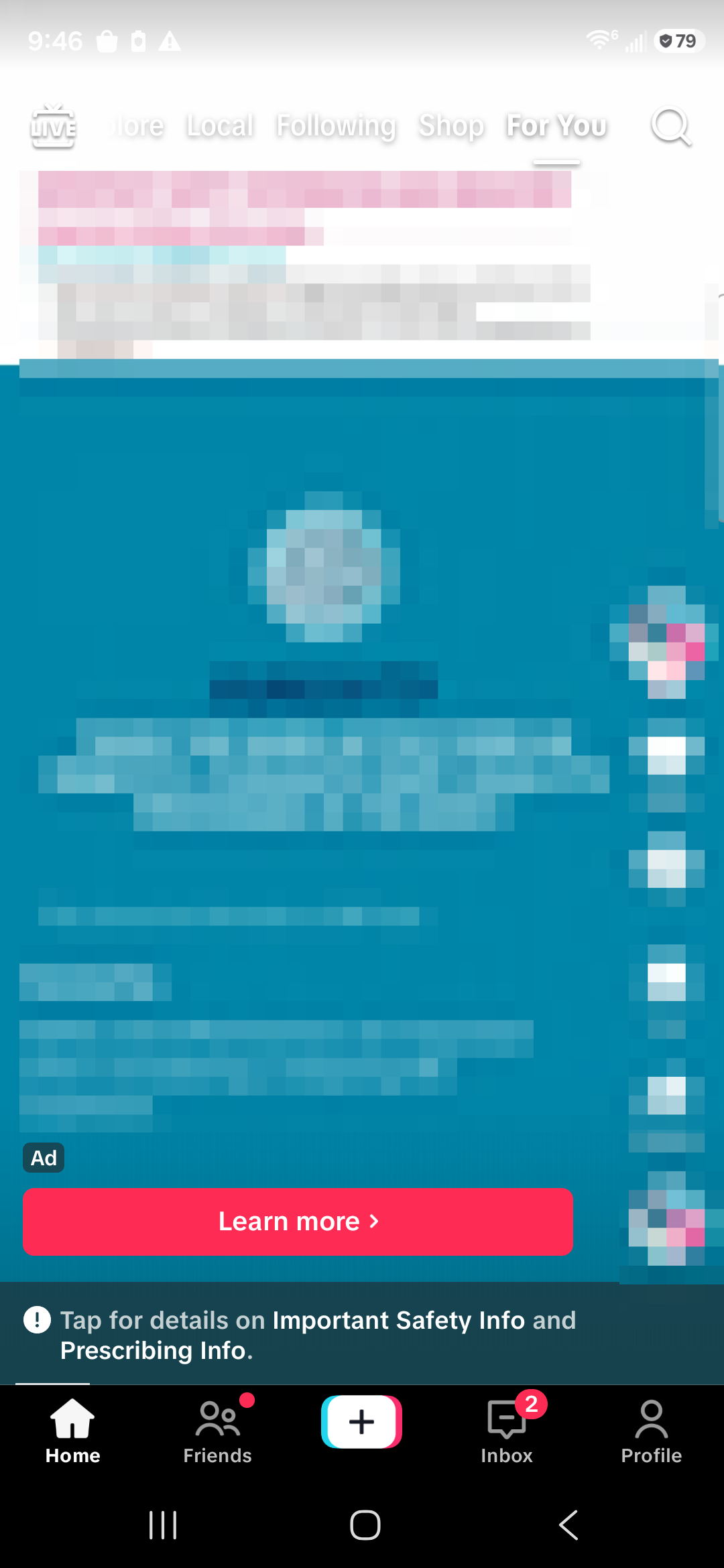}
    \includegraphics[width=0.3\linewidth]{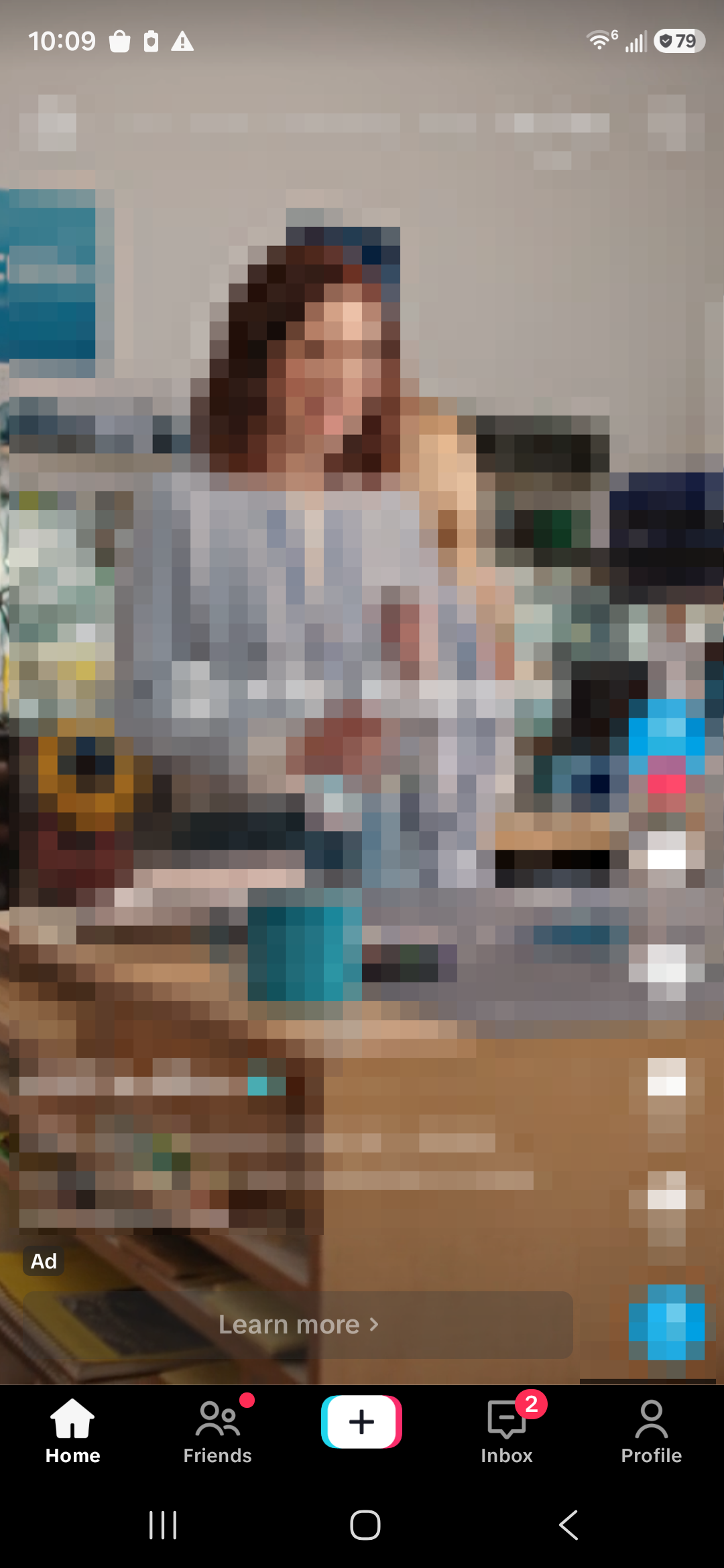}
    \includegraphics[width=0.3\linewidth]{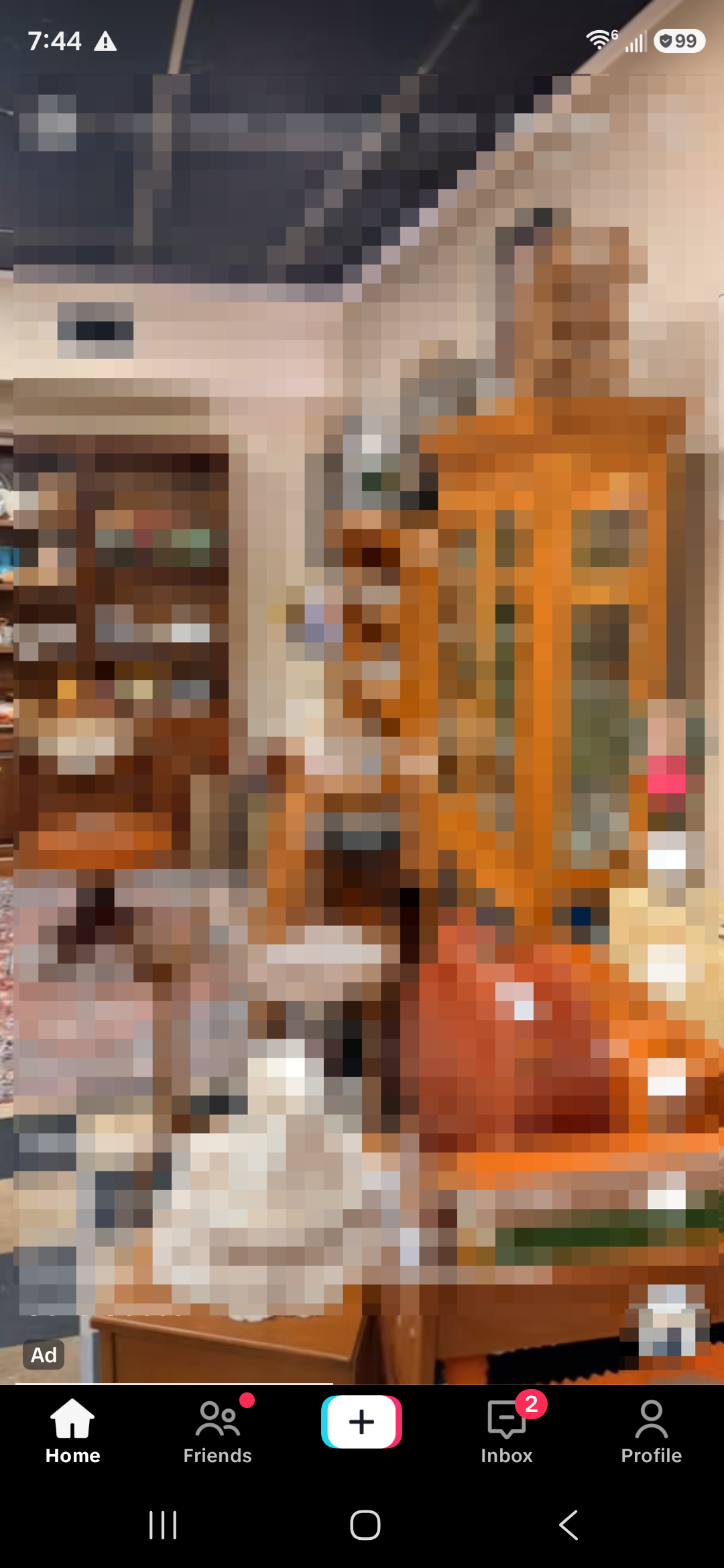}
        \caption{Red call to action button is not always present.}
        \label{fig:ad_types_all_to_action}
    \end{subfigure}
    \caption{Different ways ads are displayed on TikTok.}
    \label{fig:different_ad_visuals}
\end{figure*}

\begin{figure}
    \centering
    \includegraphics[width=0.4\linewidth]{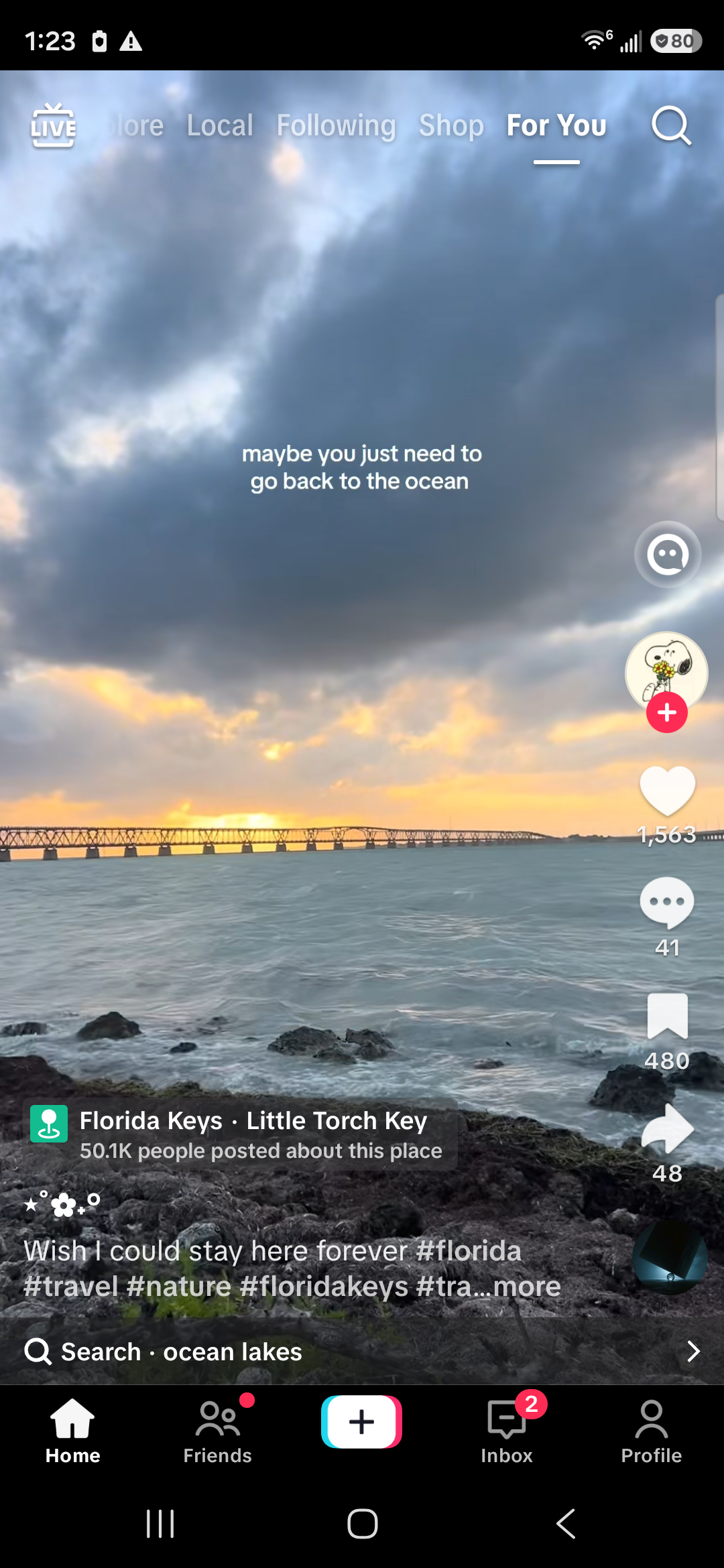}
    \caption{Location icon is not always an ad.}
    \label{fig:location_without_price}
\end{figure}

Ads can be identified on the platform using various keywords, such as: \texttt{Sponsored}, \texttt{Creator earns commission}, \texttt{Ad}, \texttt{Paid partnership}, and \texttt{Promotional content}. However, these keywords are not always present and at other times the platform might show deals using \texttt{\$+(Number)} or \texttt{(Number)+\%} to users. As shown in Figure \ref{fig:ad_multiple_tags}, the existence of these tags is not mutually exclusive. For instance, a post that has a Dollar amount could have a creator earns commission tag too, but that is not always the case. It is worth noting that the green location icon is not always an ad as it might not be followed by a Dollar or percentage off amount (see Figure \ref{fig:location_without_price}). 

\begin{figure}
    \centering
    \includegraphics[width=0.315\linewidth]{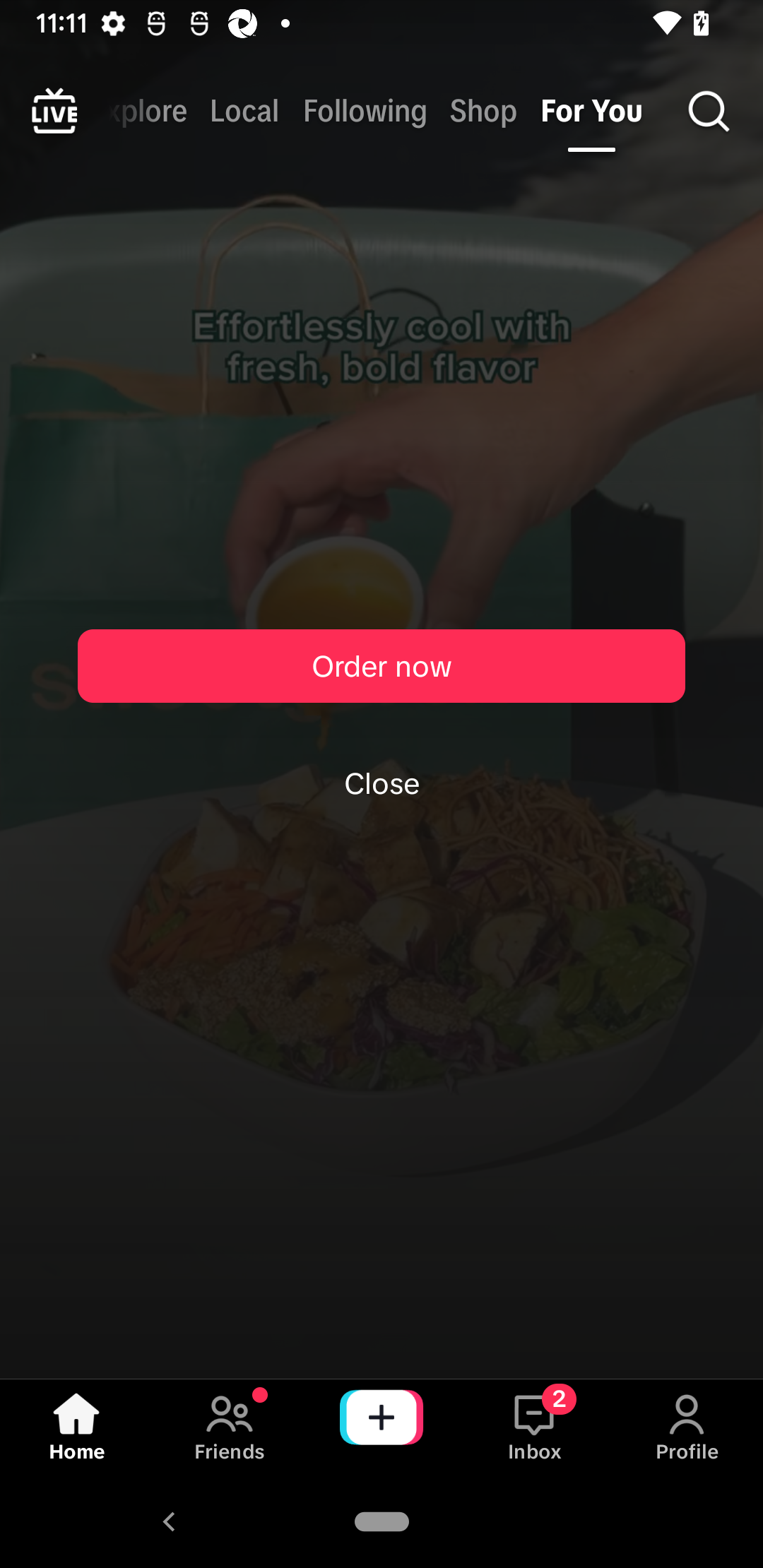}
    \includegraphics[width=0.3\linewidth]{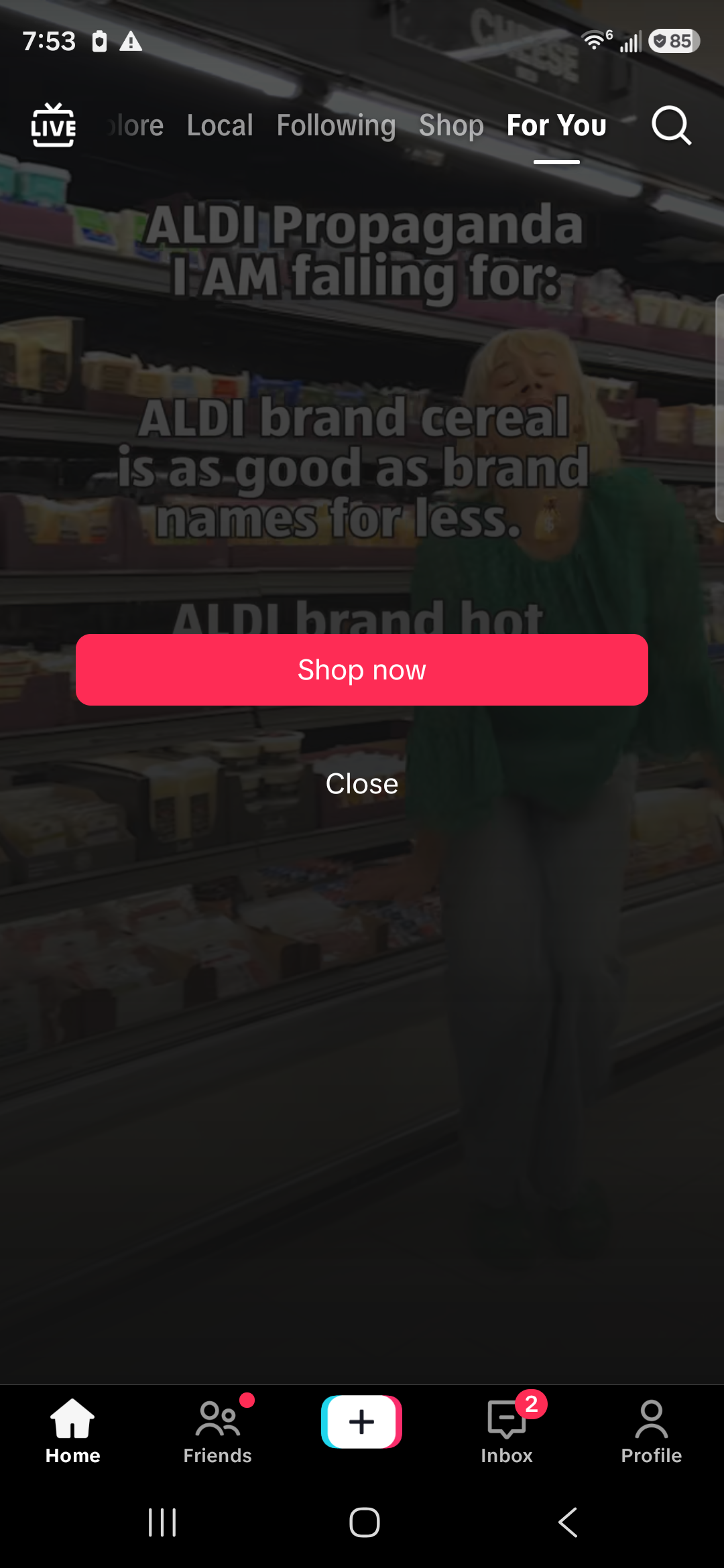}
    \includegraphics[width=0.3\linewidth]{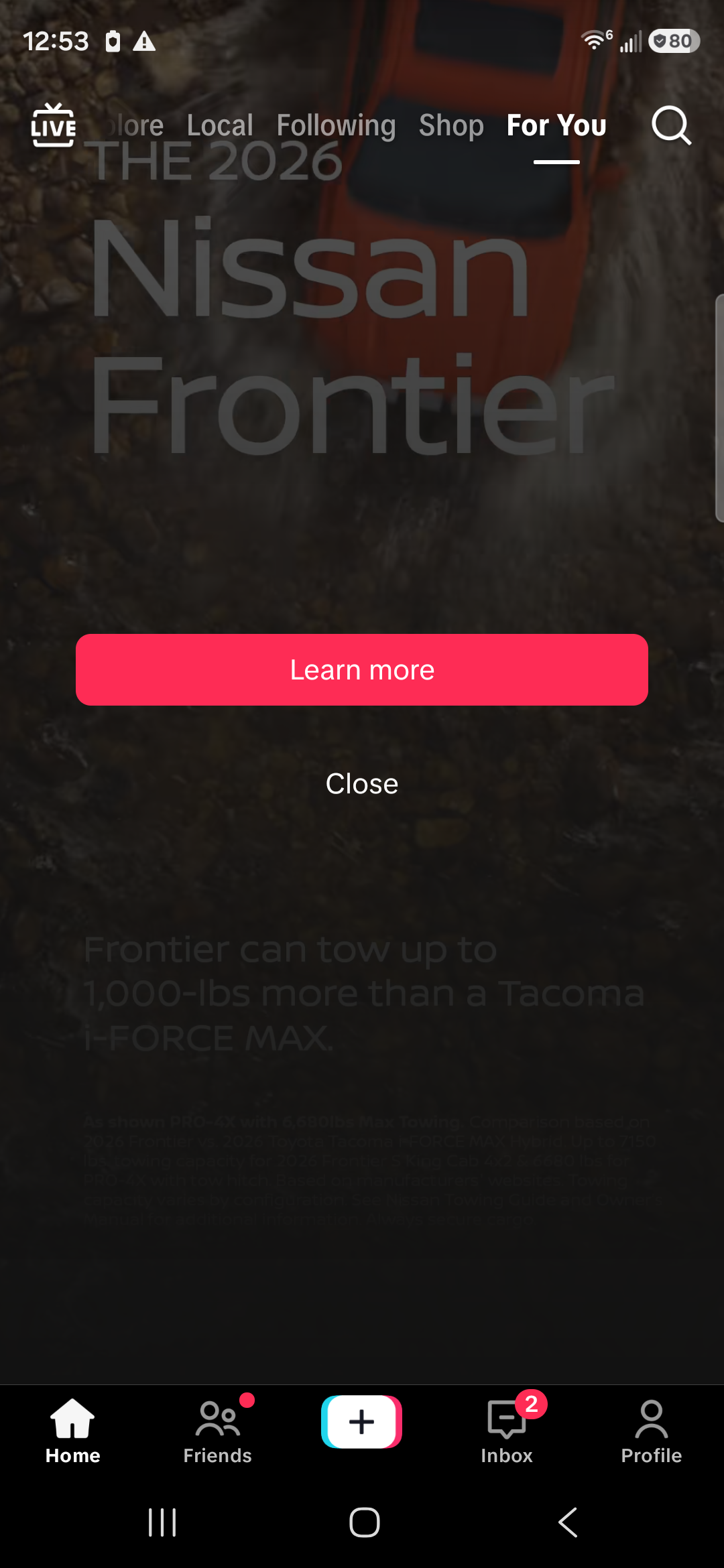}
    \caption{The red button shown in the middle of the screen, with some examples of the different texts it can have.}
    \label{fig:red_button_in_the_middle}
\end{figure}

As shown in Figure \ref{fig:ad_types_all_to_action} there is sometimes a red call-to-action button, but this button could be transparent or non-existent in some ads. Additionally, if the screenshot is taken at the end of the ad, the tags on the bottom (e.g., sponsored) will disappear and there will only be a red call to action button in the middle of the page (see Figure \ref{fig:red_button_in_the_middle}). The text on this button can be selected by the advertiser and thus can take a variety of values. Some of these values we identified in our manual review of the ads were as follows: \texttt{Shop now}, \texttt{Shop here}, \texttt{Transfer instantly}, \texttt{Get tickets now}, \texttt{Find an attorney today}, \texttt{Featured Item}, \texttt{Learn more}, \texttt{Interested}, \texttt{Sign up}, \texttt{Order now}, \texttt{Order today}, \texttt{View now}, \texttt{Limited time deal}, \texttt{Limited-time offer}, \texttt{Free shipping}, \texttt{DEALS FOR YOU}, \texttt{Book now}, \texttt{Free returns}, \texttt{Join to view hot products}, \texttt{Download}, \texttt{Download app now}, \texttt{Install now}, \texttt{File your claim here}, \texttt{Check it out}, \texttt{Go to Google Play}, \texttt{Apply now}, \texttt{Pre-order now}, \texttt{Watch now}, \texttt{Install app now}. We believe this list is not exhaustive.

\begin{figure}
    \centering
    \includegraphics[width=0.3\linewidth]{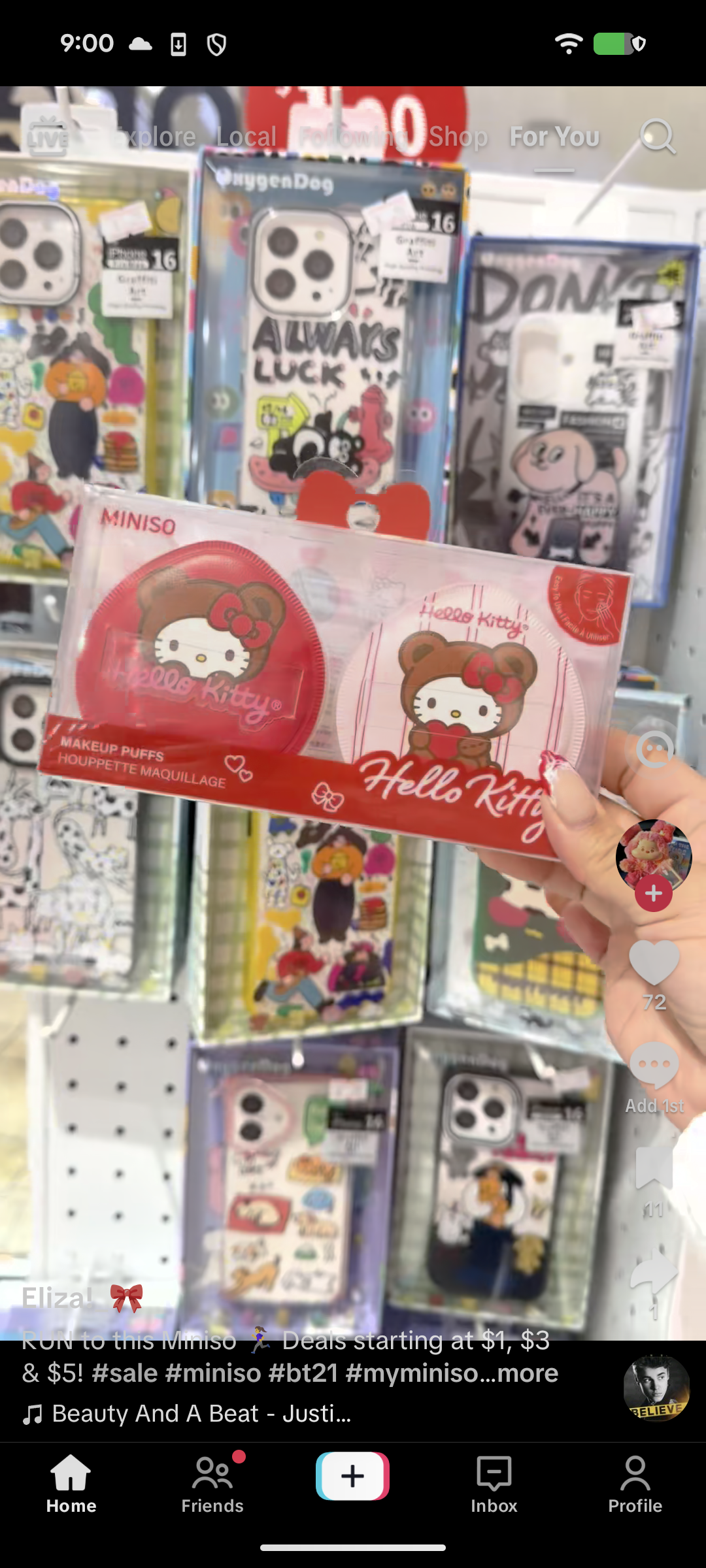}
    \includegraphics[width=0.3\linewidth]{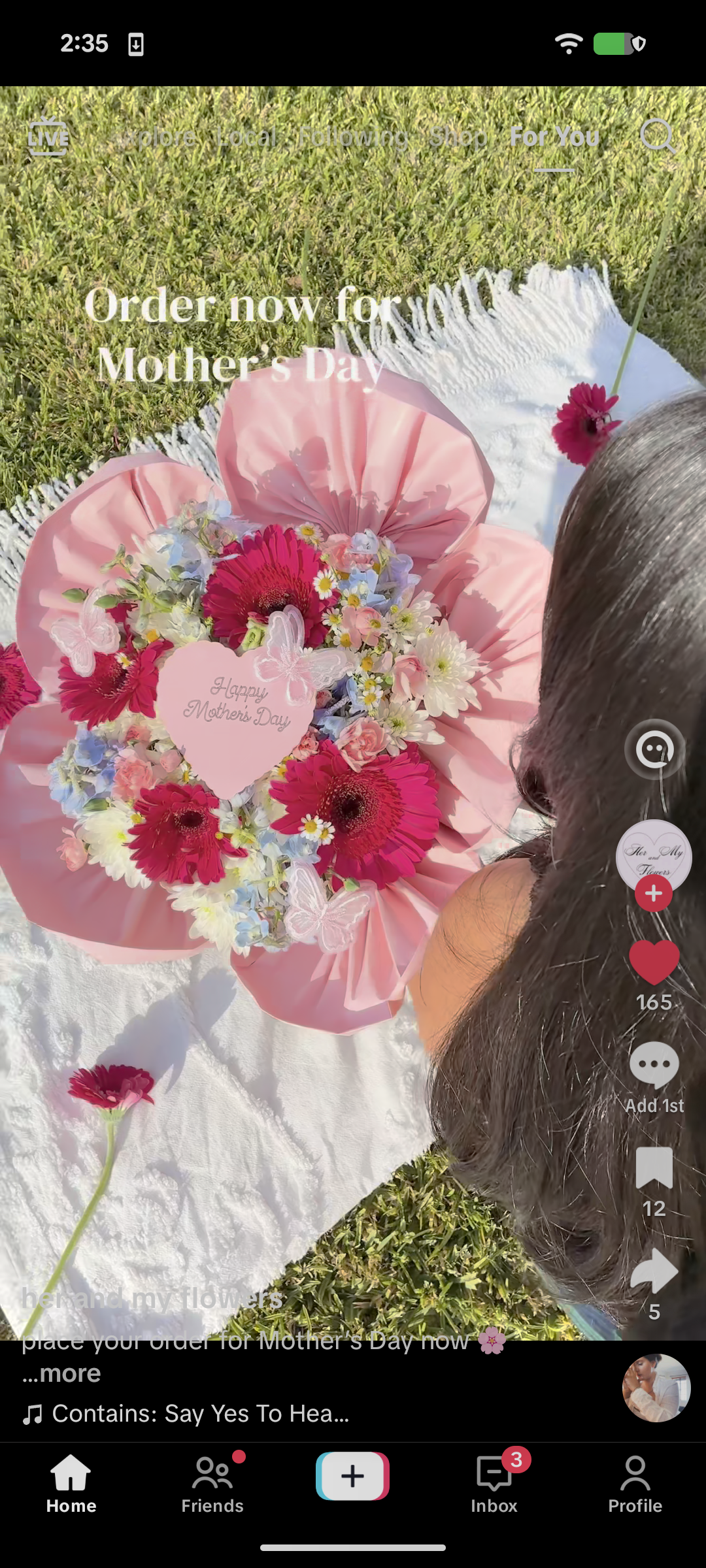}
    \includegraphics[width=0.322\linewidth]{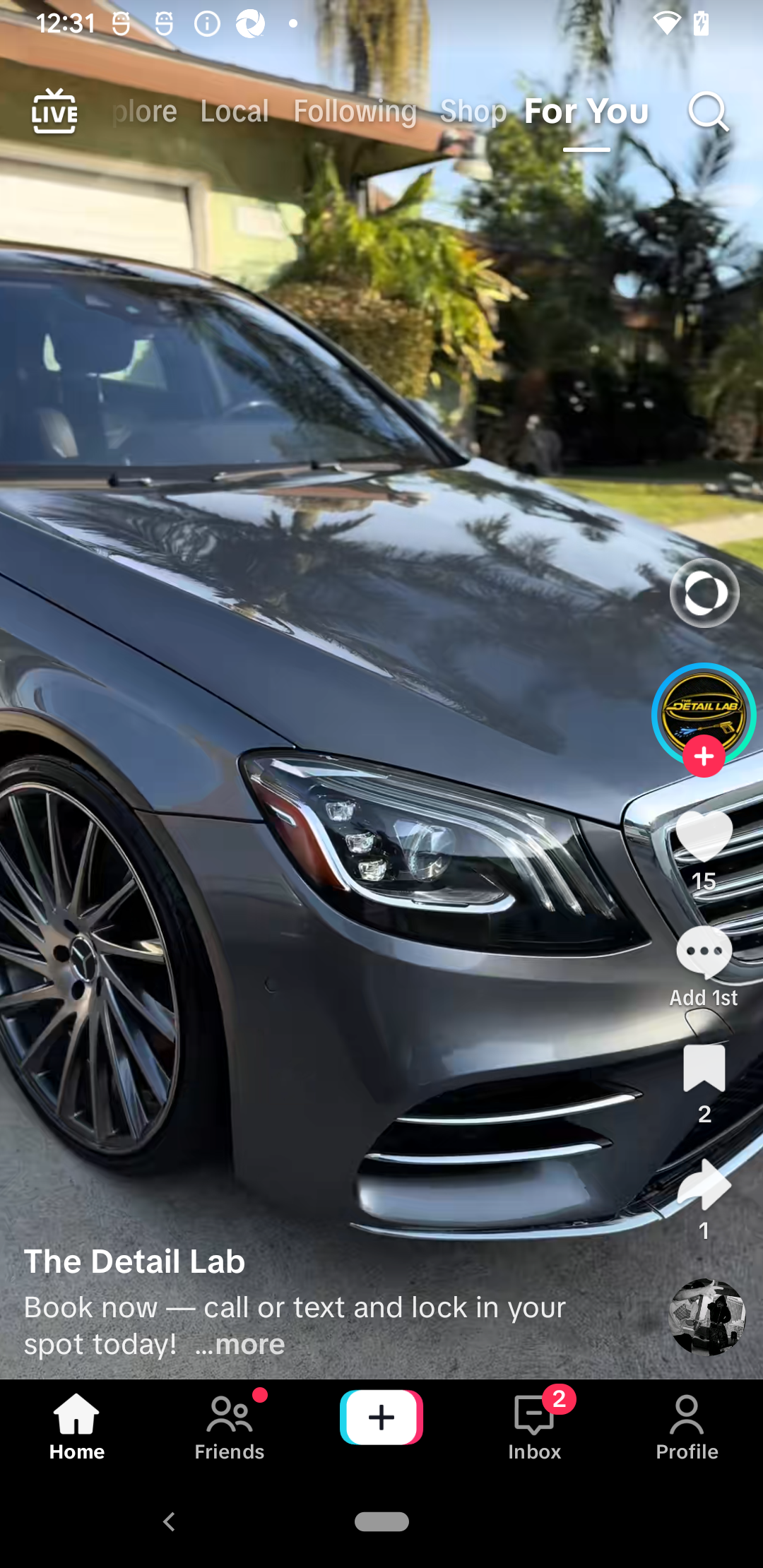}
    \caption{Posts that are ads but are not classified as such by TikTok.}
    \label{fig:ads_but_not_according_to_tiktok}
\end{figure}

It is important to note that there are some videos that are ads but are not recognized by TikTok as being ads (see Figure \ref{fig:ads_but_not_according_to_tiktok}). We try to write our ad detection code such that we don't identify these posts, because our focus is on ads officially classified by the platform. 

\subsection{B.2 Identification Process}

To detect ads we rely both on the image and the outputs of EasyOCR model. Each image is first analyzed using \texttt{opencv} to see if a red box is visible in the middle of the screen. If the box is identified, we label the screenshot as an ad and move on. Otherwise, we perform keyword matching for the different tags discussed in Appendix B.1. This code was run on all 80,169 screenshots collected through our two studies. A total of 24,255 were identified. Figure \ref{fig:tag_frequency} displays the method or tag by which ads were identified. As discussed in Appendix B.1, it is important to note that since multiple tags can be present for each ad, the high/low frequency of tags does not necessarily mean other tags were not present, it just shows how our code identified the ad. 

\begin{figure*}
    \centering
    \includegraphics[width=\linewidth]{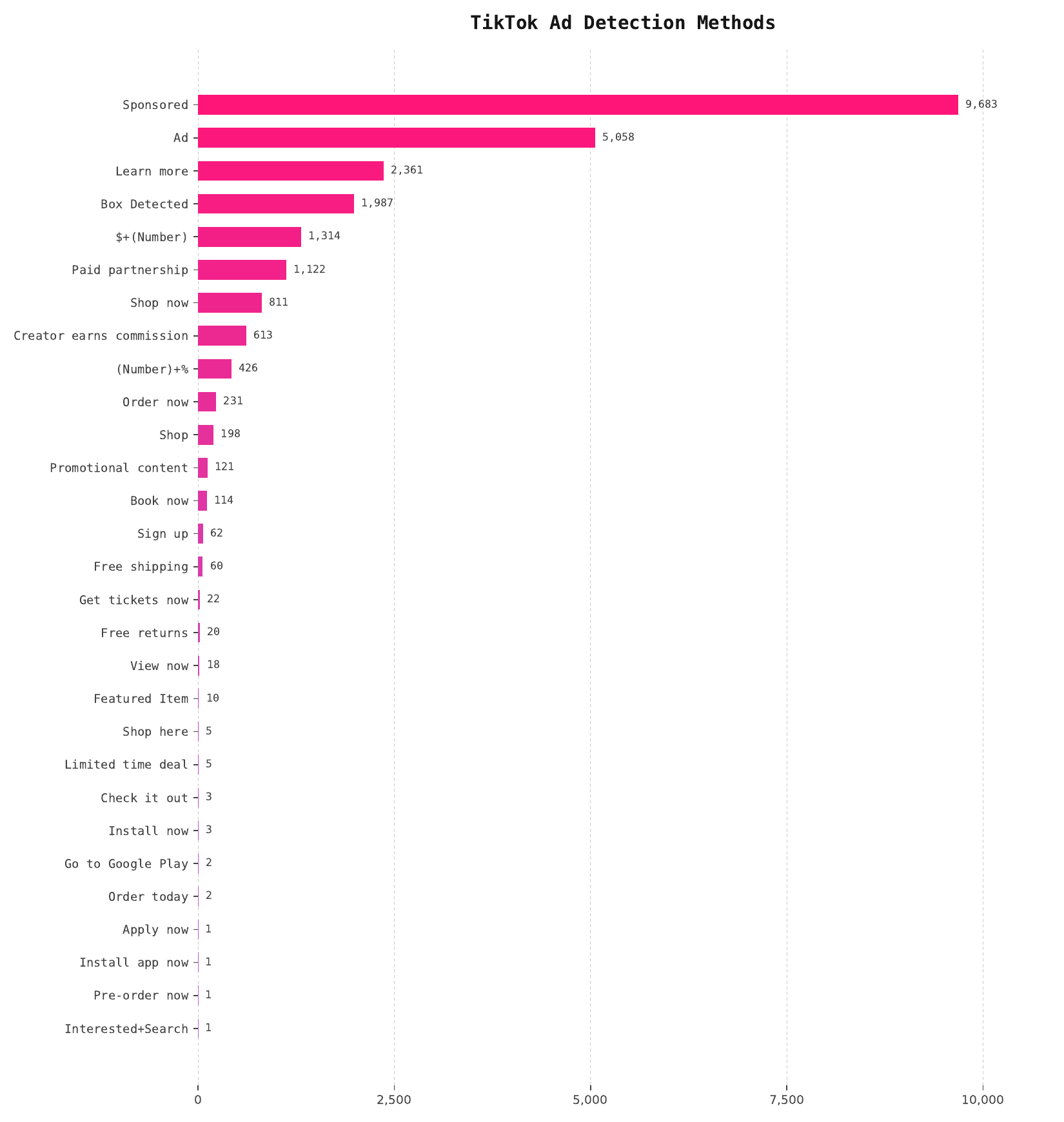}
    \caption{Frequency of tags identified through our ad detection method. The presented numbers reflect the method our code used to identify ads and do not necessarily represent the overall distribution of tags, as some ads contained multiple tags. Each ad is only counted once in this figure according to their detection method.}
    \label{fig:tag_frequency}
\end{figure*}

When matching \texttt{\$+(Number)} we only match \texttt{[f/F]rom [only] \$+(Number)}, \texttt{\$+(Number) off}, and \texttt{Get \$+(Number)}. This is done because some users might include \$ in their post but the post is not an ad identified by TikTok. The patterns for how ads use \$ were identified by one member of the research team through manual review. For \texttt{(Number)+\%} we only match \texttt{(Number)+\% off} due to similar reasons.

\subsection{Performance Evaluation}

\subsubsection{Evaluation on a True Random Sample}

To evaluate the performance of our method, we randomly sampled 500 images from our dataset of 80,169 viewed content. One member of the research team then manually reviewed these images and verified the labels. Table \ref{table:performance_of_ad_detection_true_random} shows the performance of our model.

\begin{table}[h]
\centering
\renewcommand{\arraystretch}{1.5}
\begin{tabular}{c|c|c|c}
\multicolumn{2}{c}{} & \multicolumn{2}{c}{\textbf{Actual}} \\ \cline{3-4}
\multicolumn{2}{c}{} & \textbf{Ad} & \textbf{Not Ad} \\ \cline{2-4}
\multirow{2}{*}{\rotatebox{90}{\textbf{Predicted}}} 
& \textbf{Ad} 
& \cellcolor{green!20} 142 
& \cellcolor{red!20} 0 \\ \cline{2-4}
& \textbf{Not Ad} 
& \cellcolor{red!20} 0 
& \cellcolor{green!20} 358 \\ \cline{2-4}
\end{tabular}
\caption{Performance of our ad identification method on a random sample of 500.}
\label{table:performance_of_ad_detection_true_random}
\end{table}

\subsubsection{Evaluation on a Balanced Random Sample}

To further evaluate the performance of our method, we randomly sampled 250 posts classified as ads and 250 classified as not being ads. One member of the research team then manually reviewed these images and verified the labels. Table \ref{table:performance_of_ad_detection} displays the performance of our method.

\begin{table}[h]
\centering
\renewcommand{\arraystretch}{1.5}
\begin{tabular}{c|c|c|c}
\multicolumn{2}{c}{} & \multicolumn{2}{c}{\textbf{Actual}} \\ \cline{3-4}
\multicolumn{2}{c}{} & \textbf{Ad} & \textbf{Not Ad} \\ \cline{2-4}
\multirow{2}{*}{\rotatebox{90}{\textbf{Predicted}}} 
& \textbf{Ad} 
& \cellcolor{green!20} 249 
& \cellcolor{red!20} 1 \\ \cline{2-4}
& \textbf{Not Ad} 
& \cellcolor{red!20} 1 
& \cellcolor{green!20} 249 \\ \cline{2-4}
\end{tabular}
\caption{Performance of our ad identification method on a balanced sample of 500.}
\label{table:performance_of_ad_detection}
\end{table}

\begin{figure}
    \centering
    \includegraphics[width=0.5\linewidth]{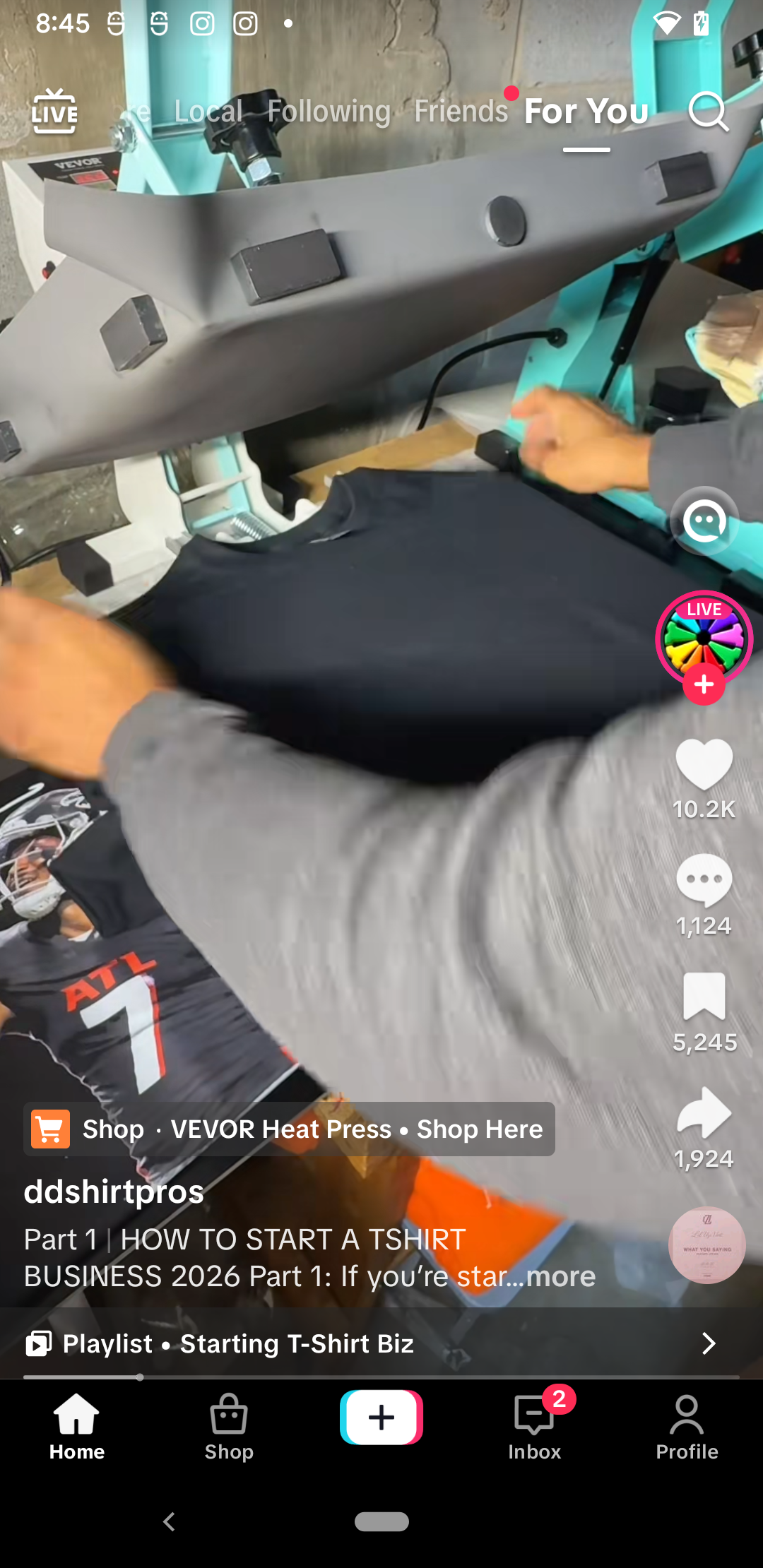}
    \caption{Example of an ad linking to TikTok shop.}
    \label{fig:tiktok_shop_eample}
\end{figure}

The one mistake that occurred, the video that was not an ad but was classified as such, was due to the word "Shop" being matched to an image that discussed being able to shop for a product but did not refer to TikTok shop. Even though we observed this issue, we decided to keep the keyword and then manually review and remove any such posts. This was because TikTok shop posts did not always have a \$ amount or \% to match them all using those methods (see Figure \ref{fig:tiktok_shop_eample}). 

\section{Analyzing Products Through Logo Detection}

We used \textit{Google Cloud Vision API}'s Logo detection function\footnote{\url{https://docs.cloud.google.com/vision/docs/detecting-logos}} to extract the logos in the 11,647 ads.

\subsection{Logos Removed From our Analysis}

\begin{figure}
    \centering
    \includegraphics[width=0.4\linewidth]{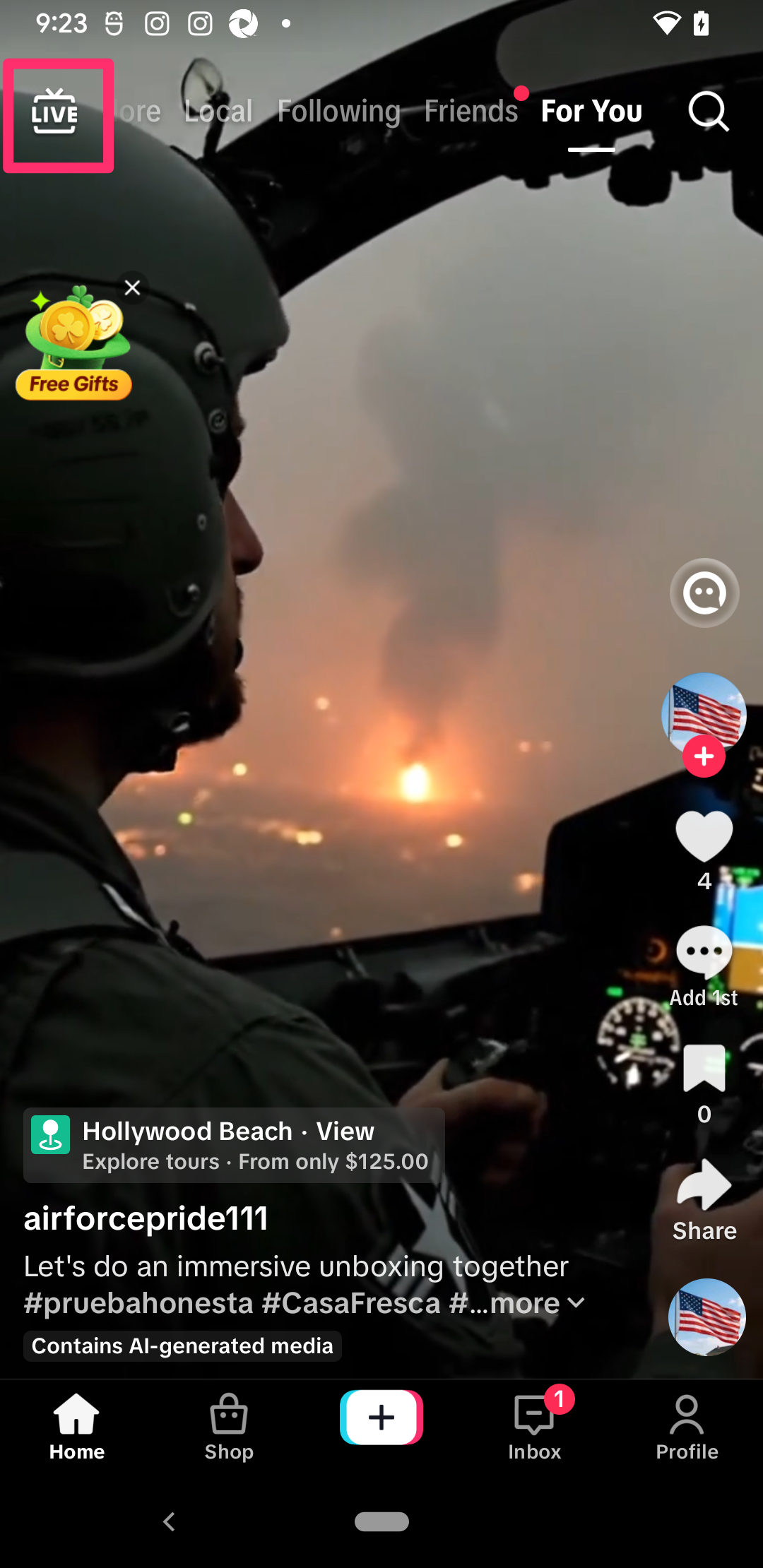}
    \caption{Part of the image that might be mistaken for the TiVo logo highlighted using a red box.}
    \label{fig:tivo_mistake}
\end{figure}

We removed TiVo from the results of our analysis, because it was mistakenly detected for 5,244 (45.0\%) of ads. We hypothesize the mistake was due to the ``Live" logo shown in Figure \ref{fig:tivo_mistake} which is similar to the Tivo logo.

\subsection{Statistics of Logo Detection Across Study 2}





Our logo detection model identified 445 unique logos across the 11,647 advertisements. We used a large language model (LLM) to assign high-level industry categories to each of the brands contained in our dataset. The model was prompted with the following instruction:

\begin{quote}
``You are given a CSV file containing brand names. For each row, determine the brand's most appropriate high-level category based on what the brand primarily sells or is known for.

Examples of categories include: Fashion \& Apparel, Beauty \& Cosmetics, Electronics, Restaurants \& Fast Food Chains, Food \& Non-Alcoholic Beverages, Alcoholic Beverages, Automotive, Insurance, Financial Services, Healthcare, Travel \& Hospitality, Software / SaaS, Social Media, Streaming Services \& Cable, Sports Teams, and Media \& Entertainment.

Instructions:
\begin{itemize}
    \item Read the CSV input.
    \item Add a new column called \texttt{category}.
    \item Output the full CSV with the new column included.
    \item Use concise, standardized category names. 
    \item Add additional categories beyond the examples when appropriate.
    \item Preserve all existing columns and row order.''
\end{itemize}
\end{quote}

We then used \texttt{Sonnet 4.6} to create initial labels for the logos. The model created 52 categories. Members of the research team then manually reviewed and fixed the categories outputted by the model. When unsure, the entire research team was consulted. One category (``Labor \& Unions") was removed because it only had 1 logo/brand, and the logo/brand was merged into ``Nonprofit \& Advocacy". Three sports-related categories were also merged into a single ``Sports" category.

The full list of 49 categories is as follows: \texttt{Advertising \& Marketing, Aerospace \& Defense, Alcoholic Beverages, Arts \& Culture, Astrology \& Spirituality, Automotive, Beauty \& Cosmetics, Collectibles \& Trading Cards, Education, Electronics, Energy \& Utilities, Environmental Services, Fashion \& Apparel, Financial Services, Flags \& Symbols, Food \& Non-Alcoholic Beverages, Gambling \& Lottery, Government \& Military, Health \& Nutrition Supplements, Healthcare, Holidays \& Observances, Home Appliances, Home Furnishings, Home Improvement, Home Security, Household \& Cleaning Products, Industrial \& Agricultural Equipment, Insurance, Logistics \& Delivery, Luxury Goods \& Jewelry, Media \& Entertainment, Music, Nonprofit \& Advocacy, Pet Products, Political Organizations, Real Estate, Religious Organizations, Restaurants \& Fast Food Chains, Retail, Social Media, Software / SaaS, Sports, Sports \& Fitness Equipment, Streaming Services \& Cable, Telecommunications, Toys \& Games, Transportation, Travel \& Hospitality, Video Games}.

\subsection{Assigning Categories to Images}

\begin{figure}
    \centering
    \includegraphics[width=0.5\linewidth]{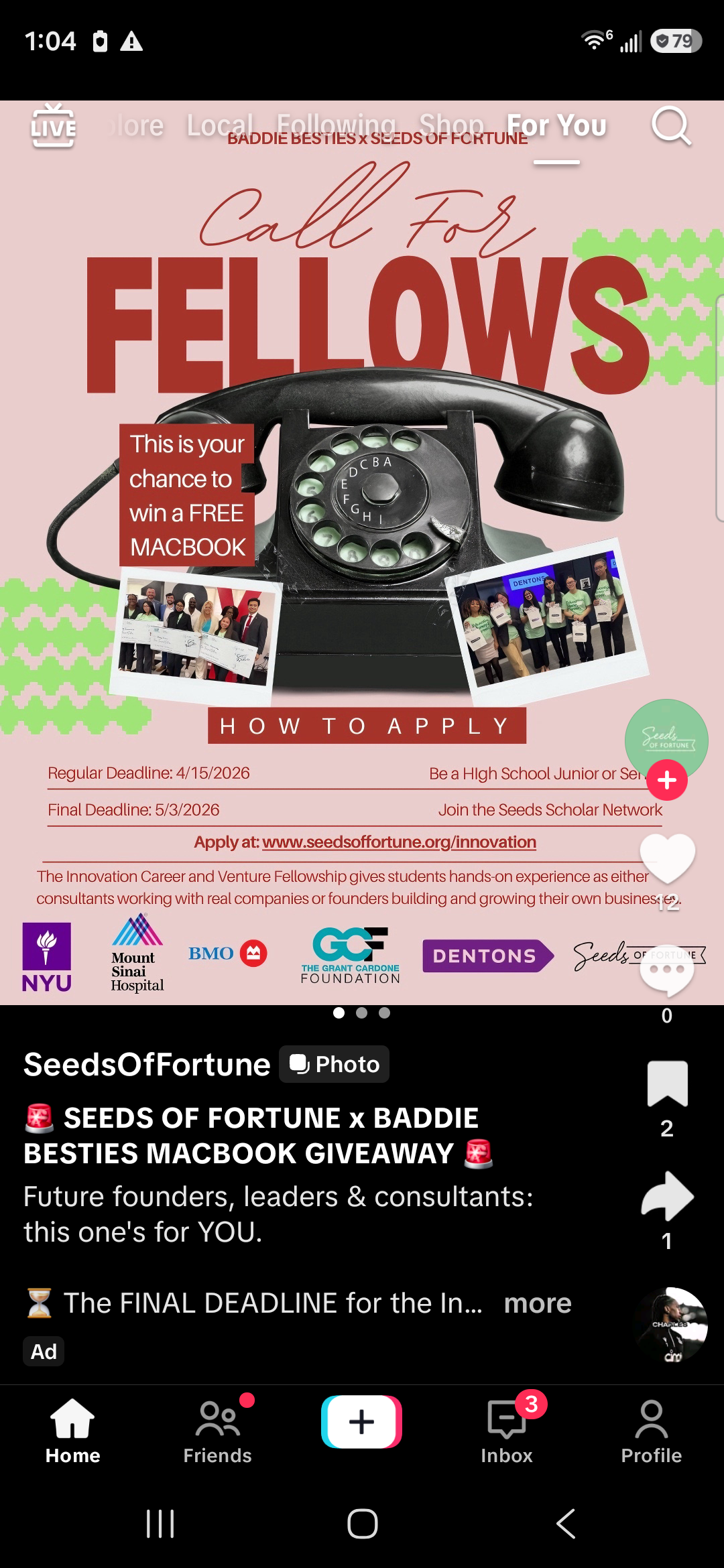}
    \caption{Example of ad with more than one logo.}
    \label{fig:ad_with_more_than_one_logo}
\end{figure}

After assigning a category to each detected logo, we assigned a single category to each advertisement image. If an advertisement contained one or more detected logos, we first mapped each logo to its manually reviewed category. We then assigned the advertisement to the category that appeared most frequently among the detected logos in that image. When multiple categories were tied for the highest frequency, we randomly selected one of the tied categories. Figure~\ref{fig:ad_with_more_than_one_logo} shows an example of an advertisement containing more than one detected logo.

\section{Categorizing Image Labels}

We used \textit{Google Cloud Vision API}'s \texttt{label\_detection} function\footnote{\url{https://docs.cloud.google.com/vision/docs/labels}} to automatically extract information about the entities within our Ad screenshots. The model outputs a description, score, and topicality for each of the recognized entities in the image. The model was run on the 11,647 screenshots identified as ads through our ad detection process. 

The model was run on screenshots after the bottom 20\% and the top 9\% of images were removed. The removal was done to try to prevent the model from outputting ``screenshot" or ``mobile device" as entities (which were observed when the full image was used as input). 

We used a second LLM prompt to map the image-based labels to the same 49 high-level categories used for logos, with an additional \texttt{Others} category used when none of the 49 categories applied. The prompt was:

\begin{quote}
``Add one CSV column named \texttt{category}. Preserve every existing row and field exactly. Return only CSV.

Allowed categories, exact spelling only: \texttt{[Full list of categories is then provided here, one category on each line. We also provide ``Other'' as an option]}. 

Decision rules:
\begin{itemize}
    \item Pick one category per row. Never invent a category.
    \item Use \texttt{Others} only when absolutely none of the non-\texttt{Others} categories can reasonably fit.
    \item If a label can plausibly fit any listed category, choose that category instead of \texttt{Others}.
    \item Do not use \texttt{Others} for any recognizable product type, media type, service, organization, team, brand, place, activity, object, industry, or consumer category.
    \item Categorize physical products by product type, not by who uses them.''
\end{itemize}
\end{quote}

The prompt also included examples and additional decision rules for common ambiguous cases, such as mapping generic people, body, emotion, animal, weather, material, and color labels to \texttt{Others} unless they clearly referred to a product, service, industry, or consumer category. The LLM mapped the 2,866 unique image-based labels into the 49 substantive categories plus \texttt{Others}. A member of the research team then manually reviewed labels assigned to \texttt{Others} to verify that they did not reasonably fit any of the 49 substantive categories.

\begin{table*}
\centering
\resizebox{\textwidth}{!}{%
\begin{tabular}{ll|ccc|ccc|ccc|ccc}
\hline
Round & Device & \multicolumn{3}{c}{Jacob} & \multicolumn{3}{c}{Emily} & \multicolumn{3}{c}{Jennifer} & \multicolumn{3}{c}{Michael} \\
 & & \textit{Like} & \textit{Comment} & \textit{Share} & \textit{Like} & \textit{Comment} & \textit{Share} & \textit{Like} & \textit{Comment} & \textit{Share} & \textit{Like} & \textit{Comment} & \textit{Share} \\
\hline
\multirow{3}{*}{1} & Pixel 3a & 94 -- 58 & 7 -- 1 & 0 -- 0 & 119 -- 63 & 8 -- 0 & 0 -- 0 & 94 -- 49 & 8 -- 5 & 0 -- 0 & 111 -- 70 & 3 -- 1 & 0 -- 0 \\
 & Pixel 8 & 72 -- 48 & 2 -- 0 & 0 -- 0 & 105 -- 63 & 3 -- 2 & 0 -- 0 & 156 -- 31 & 4 -- 1 & 0 -- 0 & 143 -- 61 & 9 -- 0 & 0 -- 0 \\
 & Galaxy S26+ & 89 -- 35 & 3 -- 1 & 0 -- 0 & 85 -- 57 & 2 -- 1 & 0 -- 0 & 132 -- 32 & 1 -- 1 & 0 -- 0 & 104 -- 46 & 6 -- 0 & 0 -- 0 \\
\hline
\multirow{3}{*}{2} & Pixel 3a & 86 -- 51 & 10 -- 2 & 0 -- 0 & 80 -- 50 & 7 -- 1 & 0 -- 0 & 96 -- 31 & 3 -- 2 & 0 -- 0 & 62 -- 36 & 9 -- 2 & 0 -- 0 \\
 & Pixel 8 & 55 -- 43 & 0 -- 2 & 0 -- 0 & 81 -- 40 & 3 -- 2 & 0 -- 0 & 89 -- 29 & 1 -- 0 & 0 -- 0 & 90 -- 58 & 2 -- 1 & 0 -- 0 \\
 & Galaxy S26+ & 100 -- 23 & 2 -- 1 & 0 -- 0 & 123 -- 44 & 7 -- 0 & 0 -- 0 & 115 -- 50 & 6 -- 1 & 0 -- 0 & 109 -- 54 & 5 -- 5 & 0 -- 0 \\
\hline
\end{tabular}
}
\caption{Number of Non-ad and ad content interacted with across our Study 2 data collection. Each cell shows (Non. Ad -- Ad) counts that have been interacted with.}
\label{tab:12phone_interactions}
\end{table*}

\section{Analyzing Interactions}

\subsection{Single Interaction Condition}

Table \ref{tab:single_interactions_counts} displays the number of content interacted with in Study 1 in which we did a single interaction per account.

\begin{table}
\centering
\begin{tabular}{llcccc}
Name & Age & Round & Like & Comment & Share \\\hline
Jacob & 26 & 1 & 72 -- 35 & 89 -- 30 & 107 -- 42 \\
 &  & 2 & 78 -- 46 & 127 -- 28 & 89 -- 39 \\[0.5em]

Emily & 26 & 1 & 97 -- 50 & 98 -- 29 & 70 -- 38 \\
 &  & 2 & 94 -- 40 & 87 -- 22 & 132 -- 73\\[0.5em]

Michael & 46 & 1 & 73 -- 33 & 71 -- 34 & 74 -- 30 \\
 &  & 2 & 78 -- 24 & 109 -- 53 & 104 -- 55\\[0.5em]

Jennifer & 46 & 1 & 88 -- 29 & 74 -- 23 & 81 -- 53 \\
 &  & 2 & 84 -- 49 & 73 -- 18 & 64 -- 54 \\
\end{tabular}
\caption{Number of Non-ad and ad content interacted with across our Study 1 data collection. Each cell shows (Non. Ad -- Ad) counts that have been interacted with.}
\label{tab:single_interactions_counts}
\end{table}

\subsection{Multiple Interaction Condition}

Table \ref{tab:12phone_interactions} displays the number of content interacted with in Study 2, in which we performed a combination of interactions within each account. Any interaction was programmed to occur with a probability of 10\%. Conditional on an interaction occurring, sharing was programmed to occur with a probability of 2\%. Given these probabilities, it is not unexpected that sharing never occurred.

\section{Study 1: Regression Model Including Age \& Gender}

Table \ref{tab:ols_single_appendix} displays a regression model for Study 1 data which includes age and gender (in addition to interaction types and total number of videos watched). \texttt{Gender} is coded 1 for the ``Jacob'' and Michael'' accounts and 0 for the ``Emily'' and ``Jennifer'' accounts. The reference group for \texttt{Age} is 26. The impact of liking and sharing on increased ad load remains statistically significant.

\begin{table}
\centering
\begin{tabular}{lrrl}
Predictor & $\beta$ Coefficient & $p$-value & \\
\hline
$\beta_0$ Intercept & -112.0654 & 0.0171 & $^{*}$ \\
$\beta_1$ \# Videos Watched & 0.3480 & 0.0000 & $^{***}$ \\
$\beta_2$ Age   & -20.5270 & 0.1673 &  \\
$\beta_3$ Gender & -8.1973 & 0.5636 &  \\
$\beta_4$ Liked  & 77.5491 & 0.0008 & $^{***}$ \\
$\beta_5$ Shared  & 135.2343 & 0.0000 & $^{***}$ \\
$\beta_6$ Commented  & 6.5105 & 0.7464 &  \\
\hline 
\multicolumn{4}{l}{%
  $n = 32$, $R^2 = 0.8947$, Adj.\ $R^2 = 0.8695$} \\
\multicolumn{4}{l}{\footnotesize Significance: $^{***}p<0.001$\ \ $^{*}p<0.05$} \\
\end{tabular}
\caption{OLS regression model predicting the number of ads based on interaction and demographic variables. Liking and sharing significantly increased ad count, while commenting had a positive but non-significant relationship.}
\label{tab:ols_single_appendix}
\end{table}

\end{document}